\documentclass[12pt]{article}

\usepackage{enumitem}
\usepackage{mathtools}

\usepackage{url}
\usepackage{amsfonts}
\usepackage{amsmath,amssymb,color}
\usepackage{graphicx}
\usepackage{enumitem}
\usepackage{multirow}
\usepackage{bm}
\usepackage{natbib}
\setcitestyle{aysep={,}}
\usepackage{verbatim}
\usepackage{float}
\usepackage[toc,page]{appendix}
\usepackage[onehalfspacing]{setspace}
\usepackage{setspace}
\usepackage{adjustbox}
\usepackage{rotating}
\usepackage{subcaption}
\usepackage{amsthm}
\usepackage{booktabs,dcolumn,caption}

\usepackage{mathabx}

\usepackage{array}

\usepackage[ruled]{algorithm2e}

\newcolumntype{L}[1]{>{\raggedright\let\newline\\\arraybackslash\hspace{0pt}}p{#1}}
\newcolumntype{C}[1]{>{\centering\let\newline\\\arraybackslash\hspace{0pt}}p{#1}}
\newcolumntype{R}[1]{>{\raggedleft\let\newline\\\arraybackslash\hspace{0pt}}p{#1}}

\newcommand{\Col}{{\mathcal{R}}}
\newcommand{\PP}{{\mathbf{P}}}

\newcommand{\constmupper}{{\kappa_1}}
\newcommand{\constmupperfinite}{{\kappa_3}}
\newcommand{\constmuppermissing}{{\kappa_5}}
\newcommand{\consttupper}{{\kappa_2}}
\newcommand{\consttupperfunction}{{\kappa_4}}
\newcommand{\consttuppermissing}{{\kappa_6}}
\newcommand{\constmlower}{{\varepsilon_1}}
\newcommand{\consttlower}{{\varepsilon_2}}

\newcommand{\ttt}{\boldsymbol \theta}

\newcommand{\aaaa}{\mathbf a}

\newcommand{\xx}{\mathbf x}
\newcommand{\ww}{\mathbf w}

\newcommand{\qq}{\mbox{$\mathbf q$}}
\newcommand{\ParaSpace}{\mathcal{S}}

\newcommand{\Real}{\mathbb{R}}
\newcommand{\Zp}{\mathbb{Z}_+}
\newcommand{\sinp}{\sin_+}

\newcommand{\argmin}{\operatornamewithlimits{arg\,min}}

\newtheorem{theorem}{Theorem}

\newtheorem{corollary}{Corollary}

\newtheorem{definition}{Definition}

\newtheorem{lemma}{Lemma}

\newtheorem{proposition}{Proposition}
\newtheorem{remark}{Remark}

\title{Structured Latent Factor Analysis for Large-scale Data: Identifiability, {Estimability, and Their} Implications}
\author{Yunxiao Chen\\
 Department of Statistics, London School of Economics and Political Science\\
 Xiaoou Li\\
 School of Statistics, University of Minnesota\\
 Siliang Zhang\\
 Shanghai Center for Mathematical Sciences, Fudan University
 }
\date{}
\begin{document}
\maketitle

\doublespacing

\begin{abstract}
Latent factor models are widely used to measure unobserved latent traits in social and behavioral sciences, including psychology, education, and marketing. When used in a confirmatory manner, design information is incorporated as zero constraints on corresponding parameters, yielding structured (confirmatory) latent factor models.
In this paper, we study how such design information affects the identifiability and the estimation of a structured latent factor model.
Insights are gained through both asymptotic and non-asymptotic analyses. Our asymptotic results are established under a regime where both the number of manifest variables and the sample size diverge, motivated by applications to large-scale data. Under this regime, we define the structural identifiability of the latent factors and establish necessary and sufficient conditions that ensure structural identifiability.  In addition, we propose an estimator which is shown to be consistent {and rate optimal} when structural identifiability holds. Finally, a non-asymptotic error bound is derived for this estimator, through which the effect of design information is further quantified. Our results shed lights on the design of large-scale measurement in education and psychology and have important implications on measurement validity and reliability.

\end{abstract}	
\noindent
KEY WORDS:
High-dimensional latent factor model, confirmatory factor analysis, identifiability of latent factors, structured low-rank matrix, large-scale psychological measurement

\section{Introduction}

Latent factor models are one of the major statistical tools for multivariate data analysis
that have received many applications in  social and behavioral sciences \citep{anderson2003introduction,bartholomew2011latent}.
Such models capture and interpret the common dependence among multiple manifest variables through the introduction of low-dimensional latent factors, where the latent factors are often given substantive interpretations. For example, in educational and psychological measurement, the manifest variables may be a test-taker's responses to test items, and the latent factors are often interpreted as his/her cognitive abilities and psychological traits, respectively. In marketing, the manifest variables may be an audience's ratings on movies, and the latent factors may be interpreted as his/her preferences on multiple characteristics of movies.

In many applications, latent factor models are used in a confirmatory manner, where design information on the relationship between manifest variables and latent factors is specified a priori.
This is translated into a zero constraint on a loading parameter of the model. We call such a model  a \textit{structured latent factor model}. For example, consider a
mathematics test consisting of $J$ items that measures students' abilities on calculus and algebra. Then a structured latent factor model with two factors may be specified to model its item response data, where the two factors may be
interpreted as the calculus and algebra factors, respectively.
The relationship between the manifest variables and the two factors is coded by a $J\times 2$ binary matrix. The $j$th row being $(1, 0)$, $(0, 1)$, and $(1, 1)$ means that the $j$th items can be solved using only calculus skill, only algebra skill, and both, respectively.

Identifiability is an important property of a structured latent factor model for ensuring the substantive interpretation of latent factors.
When the model is not identifiable, certain latent factors cannot be uniquely extracted from data and thus their substantive interpretations may not be valid.
In particular, if no design information is incorporated, then it is well known that a latent factor model is not identifiable due to rotational indeterminacy, in which case one can simultaneously rotate the latent factors and the loading matrix of the model, without changing the distribution of data. It is thus important to study the relationship between design information of manifest variables and identifiability of latent factors. 
That is, 
when is a latent factor identifiable and when is not? 
A related problem is the estimation of a structured latent factor model. Under what notion of consistency, can a latent factors be consistently estimated? In that case, what is the convergence rate? These problems will be investigated in this paper. Similar identifiability problems have been considered in the context of linear factor models \citep{anderson1956statistical,shapiro1985identifiability,grayson1994identification} and restricted latent class models \citep{xu2016identifiability,Xu2017,gu2018sufficient}.

This paper considers the identifiability and estimability of structured latent factor models.
We study this problem under  a generalized latent factor modeling framework, which includes latent factor models for continuous, categorical, and count data.
Unlike many existing works that take an empirical Bayes framework treating the latent factors as random variables, we treat them as unknown model parameters. This treatment has been taken in many works on latent factor models \citep{haberman1977maximum,holland1990sampling, bai2012statistical, owen2016bi}. Under this formulation, the identification of a latent factor becomes equivalent to the identification of the direction of a vector in $\mathbb R^{\infty}$, where each entry of the infinite-dimensional vector corresponds to a sample (e.g., a test-taker or a costumer) in a population. As will be shown in the sequel, to identify such a direction, we need an infinite number of manifest variables (e.g., measurements). Under such an asymptotic regime, we provide a necessary and sufficient condition on the design of manifest variables for the identifiability of latent factors. This condition can be very easily interpreted from a repeated measurement perspective. In the rest of the paper, this notation of identifiability is called \textit{structural identifiability}.  Comparing with existing works \citep[e.g.,][]{anderson1956statistical,shapiro1985identifiability,grayson1994identification}, the current development applies to a more general family of latent factor models, and avoids distribution assumptions on the latent variables by adopting the double asymptotic regime that the sample size $N$ and the number of manifest variables $J$ simultaneously grow to infinity.

Under the double asymptotic regime and when structural identifiability holds, we propose an estimator and establish its consistency and convergence rate. Under suitable conditions, this convergence rate is shown to be optimal through an asymptotic lower bound for this estimation problem. 
Finally, a non-asymptotic error bound is derived for this estimator, which complements the asymptotic results by quantifying the estimation accuracy under a given design with finite $N$ and $J$.
To establish these results,
we prove useful probabilistic error bounds and develop perturbation bounds on the intersection of linear subspaces which are of independent value for the theoretical analysis of low-rank matrix estimation.

%

The rest of the paper is organized as follows.
In Section~\ref{SLFA}, we introduce a generalized latent factor modeling framework, within which our research questions are formulated. In Section~\ref{sec:iden},
we discuss the structural identifiability for latent factors,
establish the relationship between structural identifiability and estimability,
and provide an estimator for which asymptotic results and a non-asymptotic error bound are established.
Further implications of our theoretical results on large-scale measurement
are provided in Section~\ref{sec:implication} and extensions of our results to more complex settings are discussed in
Section~\ref{sec:extension}.
A new  perturbation bound on linear subspaces is presented in Section~\ref{sec:proof-strategy} that is
key to our main results. Numerical results are presented in Section~\ref{sec:sim} that contain
two simulation studies and an application to a personality assessment dataset.
Finally, concluding remarks are provided in Section~\ref{sec:conc}.
The proofs of all the technical results are provided as supplementary material.

\section{Structured Latent Factor Analysis}\label{SLFA}

\subsection{Generalized Latent Factor Model}\label{subsec:model}

Suppose that there are $N$ individuals (e.g., $N$ test-takers) and $J$ manifest variables (e.g. $J$ test items). Let $Y_{ij}$ be a random variable denoting the $i$th individual's value on the $j$th manifest variable and let $y_{ij}$ be its realization.  For example, in educational tests, $Y_{ij}$s could be binary responses from the examinees, indicating whether the answers are correct or not.
We further assume that each individual $i$ is associated with a $K$-dimensional latent vector, denoted as $\ttt_i = (\theta_{i1}, ..., \theta_{iK})^{\top}$ and each manifest variable $j$ is associated with $K$ parameters $\aaaa_j = (a_{j1}, ..., a_{jK})^{\top}$. 
We give two concrete contexts. Consider an educational test of mathematics, with
$K = 3$ dimensions of ``algebra",  ``geometry", and ``calculus". Then $\theta_{i1}$, $\theta_{i2}$, and $\theta_{i3}$
represent individual $i$'s proficiency levels on algebra, geometry, and calculus, respectively.
In the measurement of  Big Five  personality factors \citep{goldberg1993structure},  $K = 5$ personality factors are considered, including ``openness to experience", ``conscientiousness", ``extraversion", ``agreeableness", and ``neuroticism". Then $\theta_{i1}$, ..., $\theta_{i5}$ represent individual $i$'s levels on the continuums of the five personality traits.
The manifest parameter $\aaaa_j$s can be understood as the regression coefficients when regressing
$Y_{ij}$s on $\ttt_i$s, $i = 1, ..., N$.
In many applications of latent factor models, especially in psychology and education, 
the estimations of $\ttt_i$s and $\aaaa_j$s are both  of interest. 

Our development is under a generalized latent factor model framework \citep{skrondal2004generalized}, which extends
the generalized linear model framework \citep{mccullagh1989generalized} to latent factor analysis.
Specifically, we assume that the distribution of $Y_{ij}$ given $\ttt_i$ and $\aaaa_j$ is a member of the exponential family with natural parameter
\begin{equation}\label{eq:inner}
m_{ij} = \aaaa_j^\top \ttt_i = a_{j1}\theta_{i1} + \cdots a_{jK}\theta_{iK},
\end{equation}
 and possibly a scale (i.e. dispersion) parameter $\phi$. More precisely, the density/probability mass function takes the form:
\begin{equation}\label{eq:model}
f(y \vert \aaaa_j, \ttt_i, \phi) = \exp\left(\frac{ym_{ij} - b(m_{ij})}{\phi} + c(y, \phi)\right),
\end{equation}
where $b(\cdot)$ and $c(\cdot)$ are pre-specified functions that depend on the member of the exponential family. Given
$\ttt_i$ and $\aaaa_j$, $i = 1, ..., N$ and $j = 1, ..., J$, we assume that all $Y_{ij}$s are independent. Consequently, the likelihood function,
in which $\ttt_i$s and $\aaaa_j$s are treated as fixed effects, can be written as
\begin{equation}\label{eq:lik}
L(\ttt_1, ..., \ttt_N, \aaaa_1, ..., \aaaa_J, \phi) = \prod_{i=1}^N\prod_{j=1}^J \exp\left(\frac{y_{ij} m_{ij} - b(m_{ij})}{\phi} + c(y_{ij}, \phi)\right).
\end{equation}
This likelihood function is known as the \textit{joint likelihood} function in the literature of latent variable models \citep{skrondal2004generalized}.
We remark that
in the existing literature of latent factor models,
there is often an intercept term indexed by $j$ in the specification of \eqref{eq:inner},
which can be easily realized under our formulation
by constraining $\theta_{i1} = 1$, for all $i = 1, 2, ..., N$. In that case, $a_{j1}$ serves as the intercept term.
This framework provides a large class of models, including linear factor models for continuous data, as well as logistic and Poisson models for multivariate binary and count data, as special cases. These special cases are listed below.

%
\begin{enumerate}
  \item Linear factor model: $$Y_{ij} \vert \ttt_i, \aaaa_j \sim N(\aaaa_j^\top \ttt_i, \sigma^2),$$ where the  scale parameter $\phi = \sigma^2$.
  \item Multidimensional Item Response Theory (MIRT) model: $$Y_{ij} \vert \ttt_i, \aaaa_j \sim Bernoulli\left(\frac{\exp(\aaaa_j^\top \ttt_i)}{1+\exp(\aaaa_j^\top \ttt_i)}\right),$$ where the scale parameter $\phi = 1$.
  \item Poisson factor model: $$Y_{ij} \vert \ttt_i, \aaaa_j \sim Poisson\left(\exp(\aaaa_j^\top \ttt_i)\right),$$ where the scale parameter $\phi = 1$.
\end{enumerate}

The analysis of this paper is based on the joint likelihood function~\eqref{eq:lik} where both $\ttt_i$s and $\aaaa_j$s are treated as fixed effects, though in the literature the person parameters $\ttt_i$ are often treated as random effects and integrated out in the likelihood function \citep{holland1990sampling}.
This fixed effect point of view allows us to straightforwardly treat the measurement problem as an estimation problem.
For ease of exposition, we assume the scale parameter is known in the rest of the paper, while pointing out that
it is straightforward to extend all the results to the case where it is unknown. 

%
\subsection{Confirmatory Structure}

In this paper, we consider a confirmatory setting where the relationship between the manifest variables and the latent factors is known a priori. Suppose that there are $J$ manifest variables and $K$ latent factors. Then the confirmatory information is recorded by a $J\times K$ matrix, denoted by $Q = (q_{jk})_{J\times K}$, whose entries take value zero or one. In particular, $q_{jk} = 0$ means that manifest variable $j$ is not directly associated with latent factor $k$. Such design information is often available in many applications of latent factor models, including in education, psychology, economics, and marketing \citep{thompson2004exploratory,reckase2009multidimensional,gatignon2003statistical}.

The design matrix $Q$ incorporates domain knowledge into the statistical analysis by imposing zero constraints on model parameters. For the generalized latent factor model, the loading parameter $a_{jk}$ is constrained to zero if $q_{jk}$ is zero.
The constraints induced by the design matrix play an important role in the identifiability and the interpretation of the latent factors.
Intuitively, suitable zero constraints on loading parameters will anchor the latent factors by preventing rotational indeterminacy,
a major problem for the identifiability of latent factor models. In the rest of this paper, we formalize this intuition by studying how the design matrix $Q$ affects the identifiability and estimability of generalized latent factor models.

\subsection{A Summary of Main Results}

In this paper, we investigate how design information given by $\qq_j$s affects the quality of measurement. This problem is tackled through both asymptotic analysis and a non-asymptotic error bound, under the generalized latent factor modeling framework.

Our asymptotic analysis focuses on the identifiability and estimability of the latent factors, under a setting where both $N$ and $J$ grow to infinity.
To define identifiability, consider a population of people where $N = \infty$
and a universe of manifest variables where $J = \infty$.
A latent factor $k$ is a hypothetical construct, defined by the person population.
More precisely, it is determined by
the individual latent factor scores of the entire person population, denoted by
$(\theta_{1k}^*, \theta_{2k}^* ...) \in \mathbb{R}^{\mathbb Z_+}$, where $\theta_{ik}^*$ denotes the true latent factor score of person $i$ on latent factor $k$ and $\mathbb{R}^{\mathbb Z_+}$ denotes the set of vectors with countably infinite real number components.
The identifiability
of the $k$th latent factor then is equivalent to the identifiability of
a vector in $\mathbb R^{\mathbb Z_+}$ under the distribution of an infinite dimensional random matrix, $\{Y_{ij}: i = 1, 2, ..., j = 1, 2, ...\}$. This setting is natural in the context of large-scale measurement where both $N$ and $J$ are large.

Under the above setting, this paper addresses three research questions.
First, how should the identifiability of latent factors be suitably formalized? Second, under what design
are the latent factors identifiable? Third, what is the relationship between the identifiability and estimability? In other words,  whether and to what extend can we recover the scores of an identifiable latent factor from data?

We further provide a non-asymptotic error bound to complement the asymptotic results. For finite $N$ and $J$, the effect of design information on the estimation of a latent factor
is reflected by a multiplying coefficient in the bound. 

%


%

\subsection{Preliminaries}\label{subsec:notation}

In this section, we fix some notations used throughout this paper.

\paragraph{Notations.}

\begin{enumerate}[label=\alph*.]
  \item $\Zp$: the set of all positive integers.
  \item $\Real^{\Zp}$: the set of vectors with countably infinite real number components.
  \item $\Real^{\Zp\times \{1, ..., K\}}$: the set of all the real matrices with countably infinite rows and $K$ columns.
  \item $\{0, 1\}^{\Zp\times \{1, ..., K\}}$: the set of all the binary matrices with countably infinite rows and $K$ columns.
  \item $\Theta$: the parameter matrix for the person population, $\Theta \in \Real^{\Zp\times \{1, ..., K\}}$.
  \item $A$: the parameter matrix for the manifest variable population, $A \in \Real^{\Zp\times \{1, ..., K\}}$.
  \item $Q$: the design matrix for the manifest variable population, $Q \in \{0, 1\}^{\Zp\times \{1, ..., K\}}$.
  \item $\mathbf 0$: the vector or matrix with all components being 0.
  \item $P_{\Theta,A}$: the probability distribution of $(Y_{ij}, i,j\in \mathbb{Z}_+)$, given person and item parameters $\Theta$ and $A$.
  \item $\mathbf v_{[1:m]}$: the first $m$ components of  a vector $\mathbf v$.
  \item $W_{[S_1,S_2]}$: the submatrix of a matrix $W$ formed by rows $S_1$ and columns $S_2$,   where $S_1, S_2 \subset \Zp$.
  \item $W_{[1:m, k]}$: the first $m$ components of the $k$-th column of a matrix $W$.
  \item $W_{[k]}$: the $k$-th column of a matrix $W$.
  \item $\Vert \mathbf v\Vert$: the Euclidian norm of a vector $\mathbf v$.
  \item $\sin\angle(\mathbf u, \mathbf v)$: the sine of the angle between two vectors,
{$$\sin\angle(\mathbf u, \mathbf v) =  \sqrt{1 - \frac{(\mathbf u^{\top} \mathbf v)^2}{\Vert \mathbf u\Vert^2 \Vert \mathbf v\Vert^2}},$$} 
        where $\mathbf u, \mathbf v\in \Real^{m}$, $\mathbf u, \mathbf v \neq \mathbf 0$. {We point out that the angles between vectors are assumed to belong to $[0,\pi]$ and $\sin\angle(\mathbf u, \mathbf v)\geq 0$ for all vectors $\mathbf u, \mathbf v$.}
  \item $\Vert W\Vert_F$: the Frobenius norm of a matrix $W = (w_{ij})_{m\times n}$, $\Vert W\Vert_F \triangleq \sqrt{\sum_{i=1}^m \sum_{j = 1}^n w_{ij}^2}$.
  \item $\Vert W\Vert_2$: the spectral norm of matrix $W$, i.e., the largest singular value of matrix.
  \item $\sigma_1(W)\geq \sigma_2(W)\geq...\geq \sigma_n(W)$: the singular values of a matrix $W\in \Real^{m\times n}$, in a descending order.
  \item $\gamma(W)$: a function mapping from $\Real^{\Zp\times n}$ to $\Real$, defined as
\begin{equation}\label{eq:gamma}
	\gamma(W) \triangleq  \liminf_{m\to\infty}\frac{\sigma_n(W_{[1:m,1:n]})}{\sqrt{m}}.
\end{equation}
  \item $|S|$: the cardinality of a set $S$.
  \item $N\wedge J$: the minimum value between $N$ and $J$.
\end{enumerate}

\section{Main Results}\label{sec:iden}

\subsection{Structural Identifiability}\label{sec:struct-ident}
We first formalize the definition of structural identifiability.
For two vectors with countably infinite components
$\mathbf w=(w_1,w_2,...)^{\top}, \mathbf z=(z_1,z_2,...)^{\top}\in \mathbb{R}^{\mathbb{Z}_+}$,
we define
\begin{equation}
	\sinp\angle(\mathbf w,\mathbf z)=\limsup_{n\to\infty}\sin\angle(\mathbf w_{[1:n]}, \mathbf z_{[1:n]}),
\end{equation}
which quantifies the angle between two vectors $\mathbf w$ and $\mathbf z$ in $\mathbb{R}^{\mathbb{Z}_+}$. In particular, we say the angle between $\mathbf w$ and $\mathbf z$ is zero when  $\sinp\angle(\mathbf w, \mathbf z)$ is zero.
\begin{definition}[Structural identifiability of a latent factor]\label{def:ident}
Consider the $k$th latent factor, where $k \in \{1, ..., K\}$,
and  a nonempty parameter space $\ParaSpace \subset \Real^{\Zp\times \{1, ..., K\}} \times \Real^{\Zp\times \{1, ..., K\}}$ for $(\Theta, A)$.
We say the $k$-th latent factor is structurally identifiable in the parameter space $\ParaSpace$ if for any $(\Theta,A), ({\Theta}',{A}')\in\ParaSpace$, $P_{\Theta,A}=P_{{\Theta}',{A}'}$ implies $\sinp\angle(\Theta_{[k]},{\Theta}'_{[k]})=0$.
\end{definition}
We point out that the parameter space $\ParaSpace$, {which will be specified later in this section,} is essentially determined by the design information $q_{jk}$s.
As will be shown shortly,
a good design imposes suitable constraints on the parameter space, which further ensures the structure identifiability of the latent factors.
This definition of identifiability avoids the consideration of the scale of the latent factor, which is not uniquely determined as the distribution of data only depends on $\{\ttt_i^{\top}\aaaa_j: i,j\in \Zp\}$.
Moreover, the sine measure is a canonical way to quantify the distance between two linear spaces that has been used in, for example, the well-known sine theorems  for matrix perturbation \citep{davis1963rotation,wedin1972perturbation}.
As will be shown in the sequel, this definition of structural identifiability naturally leads to a relationship between identifiability and estimability and has  important implications on psychological measurement.

We now characterize the structural identifiability under suitable regularity conditions.
We consider a design matrix $Q$ for the manifest variable population, where
$Q\in\{0,1\}^{\Zp\times \{1, ..., K\}}$.
	Our first regularity assumption is about the stability of the $Q$ matrix.
\begin{itemize}
	\item [A1] The limit
$$p_Q(S)=\lim_{J\to\infty}\frac{|\{j:  q_{jk} = 1, \mbox{~if~} k \in S \mbox{~and~}  q_{jk} = 0, \mbox{~if~} k \notin S, 1\leq j\leq J\}|}{J}$$ exists for any subset
$S\subset\{1,...,K\}$. In addition, $p_Q({\emptyset})=0$.
\end{itemize}

The above assumption requires that the frequency of manifest variables associated with and only with latent factors in $S$ converges to a limit proportion $p_Q(S)$.
In addition, $p_Q({\emptyset})=0$ implies that there are few irrelevant manifest variables.
{We point out that this assumption is adopted mainly to simplify the statements in Theorem~\ref{thm:identifiability} and Theorem~\ref{thm:consistency}. As discussed later, this assumption is automatically satisfied if $\qq_j$s are  generated under a stochastic design. This assumption will be further relaxed in Section~\ref{sec:a1} where a non-asymptotic error bound is established.}
We also make the following assumption on the generalized latent factor model, which is satisfied under most of the widely used models,  including the linear factor model, MIRT model, and the Poisson factor model listed above.
\begin{itemize}
		\item [A2] The natural parameter space $\{\nu: |b(\nu)|<\infty \}=\Real$.
\end{itemize}

Under the above assumptions, Theorem~\ref{thm:identifiability} provides a necessary and sufficient condition on the design matrix $Q$ for the structural identifiability of the $k$th latent factor.
This result is established within the parameter space
 $\ParaSpace_Q\subset \mathbb{R}^{\mathbb{Z}_+\times \{1, ..., K\} }\times \mathbb{R}^{\mathbb{Z}_+\times \{1, ..., K\}}$,
 \begin{equation}
 	\ParaSpace_Q= 	\ParaSpace_Q^{(1)}\times \ParaSpace_Q^{(2)}.
 \end{equation}
Here, we define
 \begin{equation}
 	\ParaSpace_Q^{(1)}=\left\{
\Theta\in \mathbb{R}^{\mathbb{Z}_+\times \{1, ..., K\} }: \|\ttt_i\|\leq C \text{ and }  \gamma(\Theta)>0
 	\right\}
 \end{equation}
 and
 \begin{equation}\label{eq:s-q-a}
\begin{split}
	\ParaSpace_Q^{(2)}=\Big\{
A\in \mathbb{R}^{\mathbb{Z}_+\times \{1, ..., K\}}:  \|\aaaa_j\|\leq C,  A_{[R_Q(S),S^c]}=\mathbf{0} \mbox{~for all~} S \subset \{1, ..., K\},\\\text{ and } \gamma(A_{[R_Q(S),S]})>0 \text{ for all } S, \mbox{ s.t. } p_Q(S)>0
 	\Big\},
\end{split}
 \end{equation}
where $C$ is a positive constant, the $\gamma$ function is defined in \eqref{eq:gamma}, and
	\begin{equation}
	R_Q(S)=\{
	j: q_{jk'} = 1, \mbox{~for all~} k' \in S \mbox{~and~}  q_{jk'} = 0, \mbox{~for all~} k' \notin S
	\}
\end{equation}
denotes the set of manifest variables that are associated {with and only with} latent factors in $S$.
Discussions on the parameter space are provided after the statement of Theorem~\ref{thm:identifiability}.

\begin{theorem}\label{thm:identifiability}
Under Assumptions A1 and A2, the $k$-th latent {factor} is structurally identifiable
in $\ParaSpace_Q$
if and only if
\begin{equation}\label{eq:condition-k}
		\{k\}=\bigcap_{k\in S, p_Q(S)>0} S,
\end{equation}
where we define $\bigcap_{k\in S, p_Q(S)>0} S=\emptyset$ if $p_Q(S)=0$ for all $S$ that contains $k$.
\end{theorem}

The following proposition guarantees that the parameter space is nontrivial.
\begin{proposition}\label{lemma:empty}
	For any $Q$ satisfying A1, $\ParaSpace_Q\neq \emptyset$.
\end{proposition}
We further remark on the parameter space $\ParaSpace_Q$.
First, $\ParaSpace_Q$ requires some regularities on each $\ttt_i$ and $\aaaa_j$ (i.e., $\|\ttt_i\|\leq C,\|\aaaa_j\|\leq C$) and the $A$-matrix satisfying the constraints imposed by $Q$ ($A_{[R_Q(S),S^c]}=\mathbf{0}$) for all $S$. It further requires that there is enough variation among people, quantified by
$\gamma(\Theta)>0$, where the $\gamma$ function is defined in \eqref{eq:gamma}. Note that this requirement is mild, in the sense that
if $\ttt_i$s are independent and identically distributed (i.i.d.) with a strictly positive definite covariance matrix, then  $\gamma(\Theta)>0$ a.s., according to the strong law of large numbers. Furthermore, $\gamma(A_{[R_Q(S),S]})>0 \text{ for } S \mbox{ satisfying } p_Q(S)>0$ requires that each type of manifest variables (categorized by $S$)
contains sufficient information if appearing frequently ($p_Q(S)>0$). Similar to the justification for $\Theta$,
$\gamma(A_{[R_Q(S),S]})>0$ can
also be justified by considering that $\aaaa_j$s are i.i.d. following a certain distribution for $j \in R_Q(S)$.

We provide an example to facilitate the understanding of Theorem~\ref{thm:identifiability}. Suppose that assumptions A1 and A2 hold.
If $K=2$ and $p_Q({\{
	1\}})= p_Q({\{1,2\}})=1/2$, then the second latent factor is not structurally identifiable, even if it is associated with infinitely many  manifest variables. In addition, having many manifest variables with a simple structure ensures the structural identifiability of a latent factor. That is, if $p_Q(\{k\}) > 0$, then the $k$th factor is structurally identifiable.

Finally, we briefly discuss Assumption A1.
The next proposition implies that if we have a stochastic design where $Q$ has i.i.d. rows, then Assumption A1 is satisfied almost surely.

\begin{proposition}\label{prop:Q-stoch}
Assume that the design matrix $Q$ has i.i.d. rows and $\qq_i\neq (0,...,0)$ for all $i=1,2,..$.
Then, Assumption A1 is satisfied almost surely. Moreover, let
\begin{equation}
	w_S=P\left(q_{1k}=1 \text{ for all } k\in S, \text{ and }q_{1k}=0 \text{ for all } k\notin S\right)
\end{equation}
for each $S\subset\{1,...,K\}$.
Then, $Q$ satisfies Assumption A1 with $p_S(Q)=w_S$ for all $S\subset\{1,...,K\}$ almost surely.
\end{proposition}

\subsection{Identifiability and Estimability}\label{sec:iden-and-est}
It is well known that for a fixed dimensional parametric model with i.i.d. observations, the
identifiability of a model parameter is necessary for the existence of a consistent estimator.
We extend this result to the infinite-dimensional parameter space under the current setting.
We start with a generalized definition for the consistency of estimating a latent factor.
An estimator given $N$ individuals and $J$ manifest variables is denoted by
$(\hat{\Theta}^{(N,J)},\hat{A}^{(N,J)})$, which only depends on $Y_{[1:N,1:J]}$ for all $N,J\in\mathbb{Z}_+$.

\begin{definition}[Consistency for estimating latent factor $k$]

The sequence of estimators
$\{
(\hat{\Theta}^{(N,J)},\hat{A}^{(N,J)}), N,J\in \mathbb{Z}_+
\}$ is said to
consistently estimate the latent factor $k$ if

\begin{equation}
	\sin\angle(\hat{\Theta}^{(N,J)}_{[k]},\Theta_{[1:N,k]})\overset{P_{\Theta,A}}{\to} 0, \text{ as } N,J\to\infty,
\end{equation}
for all $(\Theta,A)\in \ParaSpace_Q$.
\end{definition}

The next proposition establishes the necessity of the structural identifiability of a latent factor on its estimability.
\begin{proposition}\label{lemma:identifiable-estimable}
	If latent factor $k$ is not structurally identifiable in $\ParaSpace_Q$, then there does not exist a consistent estimator for latent factor $k$.
\end{proposition}

\begin{remark}
The above results of identifiability and estimability are all established under
a double asymptotic regime that both $N$ and $J$ grow to infinity, which
is suitable for large-scale applications where both $N$ and $J$ are large. It differs from the classical asymptotic setting \citep[e.g.,][]{anderson1956statistical,bing2017adaptive} where $J$ is fixed and $N$ grows to infinity.
The classical setting treats the latent factors as random effects and focuses on the identifiability of the loading parameters.
There are several reasons for adopting the double asymptotic regime. First, the double asymptotic regime allows us to directly focus on the
identifiability and the estimation of factor scores $\ttt_i$, which is of more interest in many applications (e.g, psychological and educational measurement) than the loading parameters. To consistently estimate the factor scores, naturally, we need the number of measurements $J$ to grow to infinity. {Second, having a varying $J$ allows us to take the number of manifest variables into consideration in the research design and data collection  when measuring certain traits with substantive meaning through a latent factor model.} Third, even if we only focus on the loading parameters, the classical asymptotic regime does not lead to consistency on the estimation of the loading parameters.
For example, for the MIRT model for binary data, one cannot consistently estimate the loading parameters, when $J$ is fixed and $N$ grows to infinity, unless we assume $\ttt_i$s to be i.i.d. samples from a certain parametric distribution and use this parametric assumption in the estimation procedure. Finally, this double asymptotic regime is not entirely new. In fact, this regime has been adopted in, for example, \cite{haberman1977maximum}, \cite{bai2012statistical}, and \cite{owen2016bi} among others,
for the asymptotic analysis of linear and nonlinear factor models.


%
%
%
\end{remark}

\subsection{Estimation and Its Consistency}\label{sec:consistency}

We further show that the structural identifiability and estimability are equivalent under our setting. For ease of exposition, let $Q\in \{0, 1\}^{\Zp\times \{1, ..., K\}}$ be a design matrix satisfying Assumption A1.
In addition, let $(\Theta^*,A^*)\in \ParaSpace_{Q}$ be the true parameters for the person and the manifest variable populations. We provide
an estimator $(\hat{\Theta}^{(N, J)},\hat{A}^{(N, J)})$
such that
$$
\sin \angle (\hat{\Theta}_{[k]}^{(N, J)}, \Theta_{[1:N,k]}^*)  \overset{P_{\Theta^*,A^*}}{\to} 0, ~~ N, J, \rightarrow \infty,$$
when $Q$ satisfies \eqref{eq:condition-k} which leads to
the structural identifiability of latent factor $k$
according to Theorem~\ref{thm:identifiability}.
Specifically, we consider the following estimator
\begin{equation}\label{eq:mle}
\begin{aligned}
	(\hat{\Theta}^{(N, J)},\hat{A}^{(N, J)})\in\argmin& - l\left(\ttt_1, ..., \ttt_N, \aaaa_1, ..., \aaaa_J\right),\\
s.t.~ & \|\ttt_i\|\leq C', \|\aaaa_j\|\leq C',\\
 & \aaaa_j \in \mathcal{D}_j, i = 1, ..., N, j = 1, ..., J, \\
\end{aligned}
\end{equation}
where $l(\ttt_1, ..., \ttt_N, \aaaa_1, ..., \aaaa_J)= \sum_{i=1}^N\sum_{j=1}^Jy_{ij}(\ttt_i^\top \aaaa_j) -b(\ttt_i^\top \aaaa_j)$,
$C'$ is any constant greater than $C$ in the definition of $\ParaSpace_Q$,
and $\mathcal{D}_j = \{\aaaa \in \Real^{K}: a_{jk}= 0 \mbox{~if~} q_{jk} = 0\}$ imposes the constraint on $\aaaa_j$. Note that maximizing $l\left(\ttt_1, ..., \ttt_N, \aaaa_1, ..., \aaaa_J\right)$ is equivalent to maximizing the joint likelihood \eqref{eq:lik}, due to the natural exponential family form.
%
The next theorem provides an error bound on $(\hat{\Theta}^{(N, J)},\hat{A}^{(N, J)})$.

\begin{theorem}\label{thm:consistency}

	Under assumptions A1-A2 and $(\Theta^*, A^*)\in\ParaSpace_{Q}$, there exists a constant $\constmupper$ (independent of $N$ and $J$, depending on the function $b$, the constant $C'$, and $K$) such that,
	\begin{equation}\label{eq:error-bound-m}
		\frac{1}{\sqrt{NJ}}E\left\|\hat{\Theta}^{(N, J)}(\hat{A}^{(N, J)})^{\top}-\Theta^*_{[1:N, 1:K]} A^{*\top}_{[1:J, 1:K]}\right\|_F\leq  \frac{\constmupper}{\sqrt{N\wedge J}}.
	\end{equation}
{Moreover}, if $Q$ satisfies
\eqref{eq:condition-k} and thus latent factor $k$ is structurally identifiable,
then there exists $\consttupper>0$ such that
\begin{equation}\label{eq:error-bound-dim-k}
		E \sin\angle(\Theta^*_{[1:N, k]},\hat{\Theta}_{[k]}^{(N, J)}) \leq \frac{\consttupper}{\sqrt{N\wedge J}}.
	\end{equation}
	\end{theorem}
Proposition~\ref{lemma:identifiable-estimable}, and Theorems~\ref{thm:identifiability} and \ref{thm:consistency} together imply that the structural identifiability and estimability over $\ParaSpace_Q$ are equivalent, which is
 summarized in the following corollary.
	\begin{corollary}
		Under Assumptions A1 and A2, there exists an estimator $(\hat{\Theta}^{(N, J)},\hat{A}^{(N, J)})$ such that
		$\lim_{N,J\to\infty}\sin\angle(\hat{\Theta}_{[k]}^{(N, J)},\Theta_{[1:N,k]})=0$ in $P_{\Theta,A}$ for all $(\Theta,A)\in\ParaSpace_Q$ if and only if the design matrix $Q$ satisfies \eqref{eq:condition-k}.
	\end{corollary}

\begin{remark}

The estimator \eqref{eq:mle} is closely related to the joint likelihood estimator in the literature of econometrics and psychometrics.
When $J$ is fixed and $N$ grows to infinity, this estimator is shown to be inconsistent, due to the simultaneous growth of the sample size and the parameter space \citep{neyman1948consistent,andersen1970asymptotic,ghosh1995inconsistent}. However, if $N$ and $J$ simultaneously grow to infinity, \cite{haberman1977maximum} shows that the joint maximum likelihood estimator is consistent under the Rasch model, a
simple unidimensional item response theory model.  Then under a general family of multidimensional item response theory models and under an exploratory factor analysis setting, an estimator similar to \eqref{eq:mle} is considered in \cite{chen2017joint} but without the zero constraints given by the $Q$-matrix.  As a result, the estimator obtained in \cite{chen2017joint} is rotationally indeterminant.

%

\end{remark}

	\begin{remark}
   	The error bound \eqref{eq:error-bound-m} holds even when one or more latent factors are not structurally identifiable. In particular, \eqref{eq:error-bound-m} holds when removing the constraint $\aaaa_j \in\mathcal{D}_j$ from \eqref{eq:mle}, which corresponds to the exploratory factor analysis setting where no design matrix $Q$ is pre-specified \citep[or in other words, $q_{jk} = 1$ for all $j$ and $k$; see the setting of][]{chen2017joint}. 
   In that case, the best one can achieve is to recover the linear space spanned by the column vectors of $\Theta^*$ and similarly
the linear space spanned by the column vectors of $A^*$. To make sense of such exploratory factor analysis results, one needs an additional rotation step to find an approximately sparse estimate of $A^*$ \citep{chen2017joint}.

	\end{remark}

\begin{remark}
{The proposed estimator \eqref{eq:mle} and its error bound are related to exact low-rank matrix completion \citep{bhaskar20151} and approximate low-rank matrix completion \citep[e.g.][]{candes2010matrix,davenport20141,cai2013max}, where a  bound similar to \eqref{eq:error-bound-m} can typically be derived.}
The key differences are (a)
the research on matrix completion is only interested in the estimation of $\Theta^* A^{*\top}$, while the current paper focuses on the estimation of $\Theta^*$ that is a fundamental problem of psychological measurement and
(b) our results are derived under a more general family of models.

\end{remark}

To further evaluate the efficiency of the proposed estimator,
we provide the following lower bounds. 
\begin{theorem}\label{thm:lower-bound}
Suppose that Assumptions A1 and A2 hold.
For $N, J \in \Zp$, let $\bar{M}^{(N,J)}$ be an arbitrary estimator which maps data $Y_{[1:N,1:J]}$ to $\mathbb R^{N\times J}$.
Then there exists $\constmlower>0$ and $N_0,J_0>0$ such that for $N\geq N_0, J\geq J_0$, there exists $(\Theta^*,A^*)\in\mathcal{S}_{Q}$ such that
\begin{equation}\label{eq:lower-bound-m}
	P_{\Theta^*,A^*}\left(\frac{1}{\sqrt{NJ}}\left\|\bar{M}^{(N,J)}-\Theta^*_{[1:N, 1:K]} (A^{*}_{[1:J, 1:K]})^{\top}\right\|_F\geq  \frac{\constmlower}{\sqrt{N\wedge J}}\right)\geq \frac{1}{2}.
\end{equation}
Moreover, let $\bar{\Theta}^{(N,J)}_{[k]}$ be an arbitrary estimator which maps data $Y_{[1:N,1:J]}$ to $\mathbb R^{N}$.
Then for each $A^*\in\mathcal{S}_Q^{(2)}$ defined in \eqref{eq:s-q-a}, there exists $\consttlower>0$ and $N_0,J_0>0$ such that for $N\geq N_0, J\geq J_0$, there exists $\Theta^*\in\mathcal{S}_Q^{(1)}$ such that
\begin{equation}\label{eq:lower-bound-theta}
	P_{\Theta^*,A^*}\left(\sin\angle(\bar{\Theta}^{(N, J)}_{[k]},\Theta^*_{[1:N, k]})\geq  \frac{\consttlower}{\sqrt{J}}\right)\geq \frac{1}{2}.
\end{equation}
\end{theorem}

Based on the asymptotic upper bounds and lower bounds obtained in Theorems~\ref{thm:consistency} and \ref{thm:lower-bound}, we have the following findings. First, the proposed estimator $\hat{\Theta}^{(N, J)}(\hat{A}^{(N, J)})^{\top}$ is rate optimal for the estimation of $\Theta^*_{[1:N, 1:K]} (A^{*}_{[1:J, 1:K]})^{\top}$, because the upper bound \eqref{eq:error-bound-m} matches the lower bound \eqref{eq:lower-bound-m}. Second, for estimating $\Theta^{*}_{[1:N,k]}$, the proposed estimator $\hat{\Theta}_{[k]}$ is rate optimal when
$\limsup_{N, J\rightarrow \infty} J/N  < \infty$. This assumption on the rate $J/N$ seems reasonable in many
applications of confirmatory generalized latent factor models, as $N$ is typically larger than $J$.
Finally, if $N=o(J)$, then the asymptotic upper bound \eqref{eq:error-bound-dim-k} and lower bound \eqref{eq:lower-bound-theta} do not match, in which case the proposed estimator may not be rate optimal. In particular, when $N=o(J)$, it is known that the lower bound for the estimation of $\Theta^{*}_{[1:N,k]}$ can be achieved under a linear factor model \citep[see][]{bai2012statistical}. However, this lower bound may not be achievable under a generalized latent factor model for categorical data, such as the MIRT model for binary data.
This problem is worth further investigation.

We end this section by providing an alternating minimization algorithm (Algorithm~\ref{alg:jmle}) for solving the optimization program~\eqref{eq:mle}, which is computationally efficient through our parallel computing implementation using Open Multi-Processing
\citep[OpenMP;][]{dagum1998openmp}.
{Specifically, to handle the constraints,
we adopt a projected gradient descent algorithm \citep[e.g.][]{parikh2014proximal} for solving \eqref{eq:opt1} and \eqref{eq:opt2} in each iteration, where the projections have closed-form solutions.}  Similar algorithms have been considered in other works, such as
\cite{udell2016generalized}, \cite{zhu2016personalized} and \cite{bi2017group},
for solving optimization problems with respect to low-rank matrices. Following Corollary 2 of \cite{grippo2000convergence}, we obtain Lemma~\ref{lemma:algorithm} below on the convergence property of the proposed algorithm.

\begin{lemma}\label{lemma:algorithm}
Any limit point of the sequence $\{\ttt_i^{(l)},\aaaa_j^{(l)}, i=1,...,N, j=1,...,J, l=1,2,...\}$ obtained from Algorithm~\ref{alg:jmle} is a critical point of the optimization program \eqref{eq:mle}.
\end{lemma}

\begin{algorithm}[!th]
    \SetAlgoLined
       	1. \KwIn{Data $(y_{ij}: 1\leq i\leq N, 1\leq j\leq J)$, dimension $K$, constraint parameter $C'$,
initial iteration number $l = 1$, and initial value $\ttt_i^{(0)}$ and $\aaaa_j^{(0)} \in \mathcal D_j$, $i = 1, ..., N$, $j = 1, ..., J$.}

2.  {\bf Alternating minimization:}

\For{$l = 1, 2, ...$}{
\For{$i = 1, 2, ..., N$}{
\begin{equation}\label{eq:opt1}
\ttt_i^{(l)} \in \argmin_{\Vert \ttt\Vert \leq C'} - l_i(\ttt), 
\end{equation}
  where
  $l_i(\ttt) = \sum_{j=1}^J y_{ij}(\ttt^\top \aaaa_j^{(l-1)}) -b(\ttt^\top \aaaa_j^{(l-1)})$.
}
\For{$j = 1, 2, ..., J$}{
\begin{equation}\label{eq:opt2}
   \aaaa_j^{(l)} \in \argmin_{\aaaa \in \mathcal{D}_j, \Vert \aaaa\Vert \leq C'} - \tilde l_j(\aaaa),
   \end{equation}
  where 
  $\tilde l_j(\aaaa) = \sum_{i=1}^N y_{ij}((\ttt_i^{(l)})^\top \aaaa) -b((\ttt_i^{(l)})^\top \aaaa)$.
}
}
        3. \KwOut{Iteratively perform Step 2 until convergence. Output $\hat \ttt_i = \ttt^{(L)}_i$, $\hat \aaaa_j = \aaaa_j^{(L)}$, $i = 1, ..., N$, $j = 1, ..., J$,
where $L$ is the last iteration number.}
\caption{Alternating minimization algorithm}
\label{alg:jmle}
\end{algorithm}

\subsection{A Non-asymptotic Error Bound}\label{sec:a1}

We further provide a non-asymptotic error bound as a complement to the asymptotic results. Through this error bound,  the effect of design information on the estimation of latent factors is quantified for finite $N$ and $J$, without requiring $N$ and $J$ to grow to infinity.
We introduce the following definition on collections of manifest varible types.
\begin{definition}[Feasible collection of subsets]
	For a given $k\subset\{1,...,K\}$, we say a collection of subsets of $\{1,...,K\}$, $\mathcal{A}$, is feasible  for the $k$th dimension if
	\begin{equation}
		\{k\}=\bigcap_{S\in \mathcal A} S.
	\end{equation}
\end{definition}
The concept of feasible collection is closely related to the necessary and sufficient condition \eqref{eq:condition-k} for the structural identifiability of the $k$th factor. 
Given the concept of feasible collection of manifest variable types, we then define an index as follows,
\begin{equation}\label{eq:sigmanj}
	\sigma_{N,J}=\min\left\{\frac{\max\limits_{\mathcal{A}: \mathcal{A}\text{ is feasible}}
	\min\limits_{S\in\mathcal{A}} \sigma_{|S|}( A^*_{[R_Q(S)\cap\{1,...,J\},S]} )
	}{\sqrt{J}},\frac{\sigma_K(\Theta^*_{[1:N,1:K]})}{\sqrt{N}}\right\},
\end{equation}
which serves as a measure of the signal strength on dimension $k$.
We elaborate on this index. First, for a given feasible collection $\mathcal{A}$, the first quantity in the large brackets of \eqref{eq:sigmanj}, ${\min\limits_{S\in\mathcal{A}} \sigma_{|S|}( A^*_{[R_Q(S)\cap\{1,...,J\},S]} )}/{\sqrt{J}}$, measures the amount of information contained in manifest variables associated with the set of factors $S\in\mathcal{A}$. In particular, if for any feasible collection $\mathcal{A}$, there exists an $S \in \mathcal{A}$ such that manifest variables of type $S$ do not exist, then this quantity becomes zero and thus $\sigma_{N, J} = 0$ as the second term in the brackets of \eqref{eq:sigmanj} is nonnegative. In that case, the manifest variables essentially contain no information about the $k$th dimension. Second, the second term in the brackets, ${\sigma_K(\Theta^*_{[1:N,1:K]})}/{\sqrt{N}}$, measures the co-linearity of different dimensions of $\Theta^*_{[1:N,1:K]}$.
In summary, $\sigma_{N,J}$ is a nonnegative index, with
a larger value suggesting that $\Theta^*_{[1:N,k]}$ is easier to estimate.
An error bound is then established in the next theorem.

\begin{theorem}\label{thm:finite-bound}
Suppose that Assumption A2 holds and $\|\ttt_i^*\|\leq C, \|\aaaa_j^*\|\leq C$ for all $1\leq i\leq N$ and $1\leq j\leq J$.
Then, there is a constant $\constmupperfinite$, independent of $N$ and $J$, such that
	\begin{equation}\label{eq:error-bound-m-finite}
		\frac{1}{\sqrt{NJ}}E\left\|\hat{\Theta}^{(N, J)}(\hat{A}^{(N, J)})^{\top}-\Theta^*_{[1:N, 1:K]} A^{*\top}_{[1:J, 1:K]}\right\|_F\leq  \frac{\constmupperfinite}{\sqrt{N\wedge J}}.
	\end{equation}
Moreover,  if $\sigma_{N,J}> 4\constmupperfinite^{1/2}{({N\wedge J})^{-1/2}}$,  then there exists a monotone decreasing function $\consttupperfunction(\cdot):R_+\to R_+$ (independent of $N,J$ and may depend on $C'$, $K$ and function $b$) such that for $N\geq N_0$ and $J\geq J_0$,
\begin{equation}\label{eq:error-bound-k-finite}
		E \sin\angle(\Theta^*_{[1:N, k]},\hat{\Theta}_{[k]}^{(N, J)}) \leq \frac{\consttupperfunction(\sigma_{N,J}) }{\sqrt{N\wedge J}}.
	\end{equation}
The precise forms of the constant $\constmupperfinite$ and the function $\consttupperfunction(\cdot)$ are given in the Appendix.
\end{theorem}

The above theorem does not require $N$ and $J$ to grow to infinity and thus is referred to as a non-asymptotic result.
Comparing with Theorem \ref{thm:consistency}, the assumptions of Theorem~\ref{thm:finite-bound} are weaker.
That is,
Theorem~\ref{thm:finite-bound} does not require any limit-type assumptions as required in Theorem \ref{thm:consistency}, including Assumption A1, the requirement $(\Theta^*,A^*)\in\mathcal{S}_Q$, and condition \eqref{eq:condition-k}.
Instead, it quantifies the effect of
design information on the estimation of latent factors 
under a non-asymptotic setting, only requiring the signal level $\sigma_{N,J}$ is higher than the noise level $4\constmupperfinite^{1/2}{({N\wedge J})^{-1/2}}$. According to the proposition below, this assumption can be implied by the limit-type assumptions of Theorem \ref{thm:consistency} and thus is weaker.
\begin{proposition}\label{prop:new-assump-old-assump}
If Assumption A1 and condition \eqref{eq:condition-k} are satisfied and $(\Theta^*,A^*)\in\mathcal{S}_Q$, then
	$$\liminf\limits_{N,J\to\infty}\sigma_{N,J}>0.$$
\end{proposition}

\section{Further Implications}\label{sec:implication}

In this section, we discuss the implications of the above results in the applications of generalized latent factor models to large-scale educational and psychological measurement. In such applications, each manifest variable typically corresponds to an item in a test, each
individual is a test-taker,  and both $N$ and $J$ are large. The latent factors are often interpreted as individuals' cognitive abilities, psychological traits, etc., depending on the context of the test. It is often of interest to
estimate the latent factor scores $\ttt_i$.


\subsection{On the Design of Tests.}
According to Theorems~\ref{thm:identifiability} and \ref{thm:consistency}, the key to the structural identifiability and consistent estimation of factor $k$ is
\begin{equation}\label{eq:imp}
		\{k\}=\bigcap_{k\in S, p_Q(S)>0} S,
\end{equation}
which provides insights on the measurement design. First, it implies that the ``simple structure" design advocated in psychological measurement is a safe design. Under the simple structure design, each manifest variable is associated with one and only one factor. If each latent factor $k$ is associated with many manifest variables that only measure factor $k$, or more precisely $p_Q({\{k\}}) > 0$, \eqref{eq:imp} is satisfied.

Second, our result implies that a simple structure is not necessary for a good measurement design. A latent factor can still be identified even when it is always measured together with some other factors. For example, consider the $Q$-matrix in Table~\ref{tab:Q}. Under this design,
all three factors satisfy \eqref{eq:imp} even when there is no item measuring a single latent factor.

Third, \eqref{eq:imp} is not satisfied when  there exists a $k' \neq k$ and $k'\in \bigcap_{k\in S, p_Q(S)>0} S$.
That is,
almost all manifest variables that are associated with factor $k$ are also associated with factor $k'$, in the asymptotic sense.
Consequently, one cannot distinguish factor $k$ from factor $k'$, making factor $k$ structurally unidentifiable. We point out that in this case, factor $k'$ may still be structurally identifiable, as $p_Q({\{k'\}}) > 0$ is still possible.

Finally, \eqref{eq:imp} is also not satisfied when $\bigcap_{k\in S, p_Q(S)>0} S=\emptyset$.
 It implies that the factor $k$ is not structurally identifiable when the factor is not measured by a sufficient number of manifest variables.


\begin{table}
  \centering
  \begin{tabular}{c|ccc|ccc|c}
    \hline
            & 1 & 2 & 3 & 4 & 5 & 6 & $\cdots$ \\
            \hline
            & 1 & 1 & 0 & 1 & 1 & 0 & $\cdots$\\
    $Q^\top$& 1 & 0 & 1 & 1 & 0 & 1 & $\cdots$\\
            & 0 & 1 & 1 & 0 & 1 & 1 & $\cdots$\\
    \hline
  \end{tabular}
  \caption{An example of the design matrix $Q$, which has infinite rows and $K = 3$ columns. The rows of $Q$ are given by repeating the first $3\times 3$ submatrix infinite times. }\label{tab:Q}
\end{table}

\subsection{Properties of Estimated Factor Scores}

\paragraph{A useful result.}

Let $(\Theta^*,A^*)\in \ParaSpace_{Q}$ be the true parameters for the person and the manifest variable populations. We start with a lemma connecting sine angle consistency and $L_2$ consistency.
\begin{lemma}\label{lemma:sine}
Let $\ww,\ww'\in R^n$ with $\mathbf{w}\neq \mathbf{0}$ and $\mathbf{w}'\neq\mathbf{0}$ and $c=\text{sign}(\cos\angle(\ww,\ww'))$.
Then,
\begin{equation}
\Big\|\frac{\ww}{\|\ww\|}-c\frac{\ww'}{\|\ww'\|}\Big\|^2=2-2\sqrt{1-\sin^2\angle(\ww,\ww')}.
\end{equation}
\end{lemma}
Combining Theorem~\ref{thm:consistency} and Lemma~\ref{lemma:sine}, we have the 
following corollary which
establishes a relationship between the true person parameters and their estimates.
This result is the key to the rest of the results in this section.

\begin{corollary}\label{cor:similar}
  Under Assumption A1-A2 and \eqref{eq:condition-k} is satisfied for some $k$,
  then 
there exists a sequence of random variables $c_{N,J}\in\{-1,1\}$, such that
		\begin{equation}\label{eq:cn}
			\left\|\frac{\Theta^*_{[1:N, k]}}{\|\Theta^*_{[1:N, k]}\|}-c_{N,J}\frac{\hat{\Theta}^{(N, J)}_{[k]}}{\|\hat{\Theta}^{(N, J)}_{[k]}\|}\right\|
\overset{P_{\Theta^*, A^*}}{\to} 0, ~~ N, J, \rightarrow \infty.
		\end{equation}
\end{corollary}
\begin{remark}
 Corollary~\ref{cor:similar} follows directly from \eqref{eq:error-bound-dim-k}. It provides an alternative view on how $\hat{\Theta}_{[k]}^{(N, J)}$ approximates $\Theta^*_{[1:N, k]}$. Since the likelihood function depends on $\Theta_{[1:N, 1:K]}$ and $A_{[1:J, 1:K]}$ only through $\Theta_{[1:N, 1:K]} A_{[1:J, 1:K]}^\top$, the scale of $\Theta_{[1:N, k]}$ is not identifiable even when it is structurally identifiable. This phenomenon is intrinsic to latent variable models \citep[e.g.][]{skrondal2004generalized}. Corollary~\ref{cor:similar} {states}
 that $\Theta^*_{[1:N, k]}$  and $\hat{\Theta}_{[k]}^{(N, J)}$  are close in Euclidian distance after properly normalized.
 {The}
  normalized vectors $\Theta^*_{[1:N, k]}/\Vert \Theta^*_{[1:N, k]}\Vert$ and $c_{N,J} \hat \Theta_{[k]}^{(N, J)}/\Vert \hat \Theta_{[k]}^{(N, J)}\Vert$ are both of unit length. The value of $c_{N,J}$ depends on the angle between $\Theta^*_{[1:N, k]}$ and $\hat \Theta_{[k]}^{(N, J)}$. Specifically, $c_{N,J} = 1$ if $\cos \angle(\Theta^*_{[1:N, k]}, \hat \Theta_{[k]}^{(N, J)}) > 0$ and
 $c_{N,J} = -1$ otherwise.  In practice, especially in psychological measurement,  $c_{N,J}$  can typically be determined by additional domain knowledge.
\end{remark}

\paragraph{On the distribution of person population.} In psychological measurement, the distribution of true factor scores
is typically of interest, which may provide an overview of the population on the constructs being measured.
Corollary~\ref{cor:similar} implies the following proposition on the empirical distribution of the factor scores.

\begin{proposition}\label{prop:empdist}
Suppose assumptions A1-A2 are satisfied
and furthermore \eqref{eq:condition-k} is satisfied  for factor $k$.  We normalize
$\theta_{ik}^*$ and $\hat{\theta}_{ik}^{(N, J)}$ by
\begin{equation}\label{eq:normalized-theta}
			v_i=\frac{\sqrt{N}\theta_{ik}^*}{\|\Theta^{*}_{[1:N, k]}\|} \text{ and } \hat{v}_i= \frac{c_{N,J}\sqrt{N}\hat{\theta}_{ik}^{(N, J)}}{\|\hat{\Theta}_{[k]}^{(N, J)}\|}, ~~ i=1,...,N,
\end{equation}
where $c_{N,J}$ is defined and discussed in Corollary~\ref{cor:similar}.
Let $F_{N}$ and $\hat{F}_{N,J}$ be the empirical measures of $v_1,...,v_N$ and $\hat{v}_1,...,\hat{v}_N$, respectively.
Then,
			$$Wass(F_{N},\hat{F}_{N,J}) \overset{P_{\Theta^*,A^*}}{\to}  0, ~~ N, J, \rightarrow \infty,$$		where $Wass(\cdot,\cdot)$ denotes the Wasserstein distance between two {probability measures}
			$$Wass(\mu,\nu)=\sup_{h\text{ is 1-Lipschitz}}|\int h d\mu-\int hd\nu|.$$
\end{proposition}

We point out that the normalization in \eqref{eq:normalized-theta} is reasonable. Consider a random design setting where
$\theta^*_{ik}$s are i.i.d. samples from some distribution with a finite second moment. Then $F_N$ converges weakly to the distribution of $\eta / \sqrt{E\eta^2}$, where $\eta$ is a random variable following the same distribution.
Proposition~\ref{prop:empdist} then implies that when factor $k$ is structurally identifiable and both $N$ and $J$ are large,
the empirical distribution of $\hat{\theta}_{1k}^{(N, J)}, \hat{\theta}_{2k}^{(N, J)}, ..., \hat{\theta}_{Nk}^{(N, J)}$
approximates the empirical distribution of ${\theta}^*_{1k}, {\theta}^{*}_{2k}, ..., {\theta}^{*}_{Nk}$ accurately, up to a scaling. Specifically, for any 1-Lipschitz function $h$,
$\int h(x)\hat{F}_{N, J}(dx)$ is a consistent estimator for $\int h(x)F_N(dx)$ according to the definition of Wasserstein distance. Furthermore, Corollary~\ref{cor:similar} states that under the regularity conditions, $\lim_{N\rightarrow \infty} \sum_{i=1}^N (v_i - \hat v_i)^2/N = 0$, implying that   $\sum_{i=1}^N 1_{\{(v_i - \hat v_i)^2 \geq \epsilon\}}/N = 0$, for all $\epsilon > 0$. That is, most of the $\hat v_i$s will fall into a small neighborhood of the corresponding $v_i$s.

\paragraph{On ranking consistency.} The estimated factor scores may also be used to rank individuals along a certain trait. In particular, in educational testing, the ranking provides an ordering of the students' proficiency in a certain ability (e.g., calculus, algebra, etc.). Our results also imply the validity of the ranking along a latent factor when it is structurally identifiable and $N$ and $J$ are sufficiently large. More precisely, we have the following proposition.

\begin{proposition}\label{prop:ranking}
Suppose assumptions A1-A2 are satisfied
and furthermore \eqref{eq:condition-k} is satisfied  for factor $k$.
Consider $v_i$ and $\hat v_{i}$, the normalized versions of $\theta_{ik}^*$ and $\hat \theta_{ik}^{(N, J)}$ as defined in \eqref{eq:normalized-theta}.
In addition, assume that there exists a constant $\kappa_R$ such that for any sufficiently small $\epsilon>0$ and sufficiently large $N$,
		\begin{equation}\label{eq:assump-ranking}
			\frac{\sum_{i\neq i'} I{\{\vert v_i-v_{i'} \vert \leq \epsilon \}}}{N(N-1)/2} \leq \kappa_R \epsilon.
		\end{equation}
		Then,
		\begin{equation}\label{eq:ranking}
		\frac{\tau(\mathbf v,\hat{\mathbf v})}{N(N-1)/2} \overset{P_{(\Theta^*,A^*)}}{\to}  0, ~~ N, J, \rightarrow \infty,
		\end{equation}
		where $\tau(\mathbf v,\hat{\mathbf v})=\sum_{i\neq i'}I(v_i>v_{i'},\hat{v}_i<\hat{v}_{i'})+I(v_i<v_{i'},\hat{v}_i>\hat{v}_{i'})$ is the number of inconsistent pairs according to the ranks of $\mathbf v = (v_1, ..., v_N)$ and $\hat {\mathbf v} = (\hat v_1, ..., \hat v_N)$.
\end{proposition}

We point out that \eqref{eq:assump-ranking} is a mild regularity condition on the empirical distribution $F_N$. It requires that the probability mass under $F_N$ does not concentrate in any small $\epsilon$-neighborhood, which further implies that the pairs of individuals who are difficult to distinguish along factor $k$, i.e., $(i, i')$s that $v_i$ and $v_{i'}$ are close, take only a small proportion among all the $(N-1)N/2$ pairs. In fact, it can be shown that \eqref{eq:assump-ranking} is true with probability tending to 1 as $N$ grows to infinity, when $\theta_{ik}^*$s are i.i.d. samples from a distribution with a bounded density function.
Proposition~\ref{prop:ranking} then implies that if we rank the individuals using $\hat v_i$ (assuming $c_{N,J}$ can be consistently estimated based on other information), the proportion of incorrectly ranked pairs converges to 0.
Note that  $\tau(\mathbf v,\hat{\mathbf v})$ is known as the Kendall's tau distance \citep{kendall1990rank}, a widely used measure for ranking consistency.

\paragraph{On classification consistency.} Another common practice of utilizing estimated factor scores is to classify individuals into two or more groups along a certain construct. 
For example, in an educational mastery test, it is of interest to classify examinees into ``mastery" and ``nonmastery" groups according their proficiency in a certain ability \citep{lord1980applications,bartroff2008modern}. In measuring psychopathology, it is common to classify respondents into ``diseased" and ``non-diseased" groups based on a mental health disorder. We justify the validity of making classification based on the estimated factor score.

\begin{proposition}\label{prop:classify}
Suppose assumptions A1-A2 are satisfied
and furthermore \eqref{eq:condition-k} is satisfied  for factor $k$.
Consider $v_i$ and $\hat v_{i}$, the normalized versions of $\theta_{ik}^*$ and $\hat \theta_{ik}^{(N, J)}$ as defined in \eqref{eq:normalized-theta}.
Let $\tau_-<\tau_+$ be the classification thresholds, then
\begin{equation}\label{eq:classificatin}
\frac{\sum_{i=1}^N I{\left\{\hat{v}_i\geq \tau_+, v_i\leq  \tau_-  \right\}} + {I}{\left\{\hat{v}_i\leq \tau_-, v_i \geq \tau_+ \right\}}}{N} \overset{P_{\Theta^*,A^*}}{\to}  0, ~~ N, J, \rightarrow \infty,.
\end{equation}
\end{proposition}

Considering two pre-specified thresholds $\tau_-$ and $\tau_+$ is the well-known indifference zone formulation of educational mastery test \cite[e.g.][]{bartroff2008modern}. In that context, examinees with $v_i \geq \tau_+ $ are classified into the ``mastery" group and those with $v_i\leq \tau_-$ are classified into the ``nonmastery" group.
The interval  $(\tau_-, \tau_+)$ is known as the indifference zone, within which no decision is made.
Proposition~\ref{prop:classify} then implies that when factor $k$ is structurally identifiable, the classification error tends to 0 as both $N$ and $J$ grow to infinity.

\section{Extensions}\label{sec:extension}

\subsection{Generalized Latent Factor Models with Intercepts}
As mentioned in Section~\ref{subsec:model}, intercepts can be easily incorporated into the generalized latent factor model by restricting $\theta_{i1}=1$. Then, $a_{j1}$s are the intercept parameters and $q_{j1}=1$ for all $j$.
Consequently,  for any $S$ satisfying $p_Q(S) > 0$, $1\in S$ and thus
the latent factors 2-$K$ are not structurally identifiable according to Theorem~\ref{thm:identifiability}.
Interestingly, these factors are still structurally identifiable if we
restrict to the following parameter space
\begin{equation}
\begin{split}
	\ParaSpace_{Q,-}
	=
	\Big\{&
	(\Theta,A)\in  \ParaSpace_Q:
\lim_{N\to\infty}\frac{1}{N}\mathbf{1}_{N}^{\top}\Theta_{[1:N,m]}=0 \text{ for } m\geq 2, \text{ and } \theta_{i1}=1 \text{ for } i\in\Zp
	\Big\},
\end{split}
\end{equation}
which requires that $\Theta_{[k]}$ and $\Theta_{[1]}$ are asymptotically orthogonal, for all $k \geq 2$.

\begin{proposition}\label{thm:intercept}
Under Assumptions A1-A2, and assuming that $q_{j1}=1$ for all $j\in\Zp$ and $K\geq 2$,
then
	{the $k$th latent {factor} is structurally identifiable in $\ParaSpace_{Q,-}$ if and only if
	\begin{equation}\label{eq:condition-k-intercept}
		\{1, k\}=\bigcap_{k\in S, p_Q(S)>0} S,
	\end{equation}
	for $k\geq 2$.
	}
\end{proposition}
The next proposition guarantees that $\ParaSpace_{Q,-}$ is also non-empty.
\begin{proposition}\label{lemma:empty-intercept}
	For all $Q$ satisfying A1 and $q_{j1}=1$ for all $j\in\Zp$, and in addition $C>1$, then $\ParaSpace_{Q,-}\neq\emptyset$.
\end{proposition}

	\begin{remark}
		When having intercepts in the model, similar results of consistency {and a non-asymptotic error bound} can be established for the estimator
\begin{equation}\label{eq:mle2}
\begin{aligned}
	(\hat{\Theta}^{(N, J)},\hat{A}^{(N, J)})\in\argmin& - l\left(\ttt_1, ..., \ttt_N, \aaaa_1, ..., \aaaa_J\right),\\
s.t.~ & \|\ttt_i\|\leq C', \|\aaaa_j\|\leq C', \aaaa_j \in \mathcal{D}_j, \\
 & \theta_{i1} = 1, \sum_{i'=1}^N \theta_{i'k} = 0,\\
 & i = 1, ..., N, j = 1, ..., J, k = 2, ..., K.
\end{aligned}
\end{equation}
\end{remark}


\subsection{Extension to Missing Values}

Our estimator can also handle missing data which are often encountered in practice.
Let $\Omega = (\omega_{ij})_{N\times J}$ be the indicator matrix of nonmissing values, where
$\omega_{ij} = 1$ if $Y_{ij}$ is observed and $\omega_{ij} = 0$ if $Y_{ij}$ is missing. When data are completely missing at random, the joint likelihood function becomes
$$L^\Omega(\ttt_1, ..., \ttt_N, \aaaa_1, ..., \aaaa_J, \phi) = \prod_{i,j: \omega_{ij} = 1} \exp\left(\frac{y_{ij} m_{ij} - b(m_{ij})}{\phi} + c(y_{ij}, \phi)\right)$$
and our estimator becomes
\begin{equation}\label{eq:mle3}
\begin{aligned}
	(\hat{\Theta}^{(N, J)},\hat{A}^{(N, J)})\in \argmin& - l^\Omega\left(\ttt_1, ..., \ttt_N, \aaaa_1, ..., \aaaa_J\right),\\
s.t.~ & \|\ttt_i\|\leq C', \|\aaaa_j\|\leq C',\\
 & \aaaa_j \in \mathcal{D}_j, i = 1, ..., N, j = 1, ..., J, \\
\end{aligned}
\end{equation}
where $l^{\Omega}(\Theta A^{\top})= \sum_{i, j: \omega_{ij} = 1} y_{ij}m_{ij}-b(m_{ij})$.
Moreover, results similar to Theorem~\ref{thm:consistency} can be established even when having missing data.
Specifically, we assume
\begin{enumerate}
  \item[A3] $\omega_{ij}$s in $\Omega$ are independent and identically distributed Bernoulli random variables with
$$P(\omega_{ij} = 1) = \frac{n}{NJ}.$$
\end{enumerate}
This assumption implies that data are completely missing at random and only about $n$ entries of $(Y_{ij})_{N\times J}$
are observed. We have the following result.
\begin{proposition}\label{prop:missing}

Under assumptions A1-A3 and $(\Theta^*, A^*)\in\ParaSpace_{Q}$, there exists $\constmuppermissing>0$ such that,
	\begin{equation}\label{eq:error-bound-m-missing}
\begin{aligned}
		&\frac{1}{\sqrt{NJ}} E\left(\|\hat{\Theta}^{(N, J)}(\hat{A}^{(N, J)})^{\top}-\Theta^*_{[1:N, 1:K]} (A_{[1:N, 1:K]}^{*})^\top\|_F\right)\\
\leq &\constmuppermissing\max\left\{
	\sqrt{\frac{N\vee J}{n}},\frac{(NJ)^{1/2}}{n^{3/4}}
	\right\}.
\end{aligned}
	\end{equation}
Moreover, if $Q$  satisfies
\eqref{eq:condition-k} and thus latent factor $k$ is structurally identifiable,
then  there exists a constant $\consttuppermissing$, such that
	\begin{equation}\label{eq:error-bound-dim-k-missing}
		E \sin\angle(\Theta^*_{[1:N, k]},\hat{\Theta}_{[k]}^{(N, J)}) \leq \consttuppermissing \max\left\{
	\sqrt{\frac{N\vee J}{n}},\frac{(NJ)^{1/2}}{n^{3/4}}
	\right\}.
	\end{equation}
\end{proposition}
\begin{remark}
Results similar to \eqref{eq:error-bound-m-missing} have also
been derived in the matrix completion literature \citep[e.g.][]{candes2010matrix,davenport20141,cai2013max,bhaskar20151} under specific statistical models with an underlying low rank structure.
Proposition~\ref{prop:missing} extends the existing results on matrix completion to a generalized latent factor model.
\end{remark}


\section{A Useful Perturbation Bound on Linear Subspace}\label{sec:proof-strategy}
The standard approach  (see, e.g., \cite{davenport20141}) for bounding the error of the maximum likelihood estimator is by making use of the strong/weak  convexity of the log-likelihood function.
However, in the generalized latent factor model, the log-likelihood function is not convex in $(\Theta_{[1:N,1:K]},A_{[1:J,1:K]})$. Thus, the standard approach is not applicable for proving \eqref{eq:error-bound-dim-k} in Theorem~\ref{thm:consistency} and similar results on the estimation of $\Theta_{[1:N,k]}$ in Theorem~\ref{thm:finite-bound} and Proposition~\ref{prop:missing}.

New technical tools are developed to handle this problem. Specifically, there are two major steps for proving Theorem~\ref{thm:consistency}.
In the first step, we establish \eqref{eq:error-bound-m}, which bounds the error on the estimation of $\Theta_{[1:N, 1:K]}^* A^{*\top}_{[1:J, 1:K]}$ as a whole. This step extends the error bound for exact low-rank matrix completion \citep{bhaskar20151} under the generalized latent variable model setting. A small  estimation error of $\Theta_{[1:N, 1:K]}^* A^{*\top}_{[1:J, 1:K]}$ implies that the linear space spanned by the column vectors of $\Theta_{[1:N, 1:K]}^*$ can be recovered up to a small perturbation.
In the second step, given the result from the first step and design information, we show that the direction of the vector $\Theta_{[1:N, k]}^*$ can be recovered up to a small perturbation. This step is technically challenging and is tackled by a new perturbation bound for the intersection of linear spaces. This perturbation bound, as introduced below, may be of independent value for the theoretical analysis of low-rank matrix estimation.

Let $\Col(W)$ denote the column space of a matrix $W$. \sloppy
Under the conditions of Theorem~\ref{thm:consistency}, the result of \eqref{eq:error-bound-m}
combined with the Davis-Kahan-Wedin sine theorem \citep[see e.g.][]{stewart1990matrix} allows us to bound
$\sin\angle(\Col(\Theta^*_{[1:N, S]}),\Col(\hat{\Theta}^{(N, J)}_{[S]}))$, for any $S$ satisfying $p_Q(S) > 0$,
where $\angle(L,M)$ denotes the largest principal angle between two linear spaces $L$ and $M$, i.e.,
$\sin \angle(L,M) = \max_{\mathbf u\in M, \mathbf u\neq 0} \min_{\mathbf v\in L, \mathbf v\neq \mathbf 0} \sin \angle(\mathbf u,\mathbf v).$
Our strategy is to bound $$ \sin\angle(\Theta^*_{[1:N, k]},\hat{\Theta}_{[k]}^{(N, J)}) = \sin\angle(\Col(\Theta^*_{[1:N, k]}),\Col(\hat{\Theta}_{[k]}^{(N, J)}))$$ by $\sin\angle(\Col(\Theta^*_{[1:N, 1:K]}),\Col(\hat{\Theta}^{(N, J)}))$ under the assumptions of Theorem~\ref{thm:consistency}.
Note that $\Col(\Theta^*_{[1:N, k]}) = \bigcap_{k\in S, p_Q(S) > 0} \Col(\Theta^*_{[1:N, S]})$ and similarly
$\Col(\hat{\Theta}_{[k]}^{(N, J)}) = \bigcap_{k\in S, p_Q(S) > 0} \Col(\hat{\Theta}_{[S]}^{(N, J)})$.
Consequently, it remains to show that
if the linear spaces are perturbed slightly,
then their intersection does not change much.
To this end, we establish a new perturbation bound on the intersection of general linear spaces in the next proposition.

\begin{proposition}[Perturbation bound for intersection of linear spaces]\label{prop:perturb}
Let $L$, $M$, $L'$, $M'$ be linear subspaces of a finite dimensional vector space. Then,
\begin{equation}\label{eq:perturbation-bound}
	\|\PP_{L'\cap M'}-\PP_{L\cap M}\|\leq  8 \max\{
      \alpha(\theta_{\min,+}(L,M)),\alpha(\theta_{\min,+}( L', M'))
      \}(\|\PP_L-\PP_{L'}\|+\|\PP_{M}-\PP_{M'}\|),
\end{equation}
where  we define $\theta_{\min,+}(L,M)$ as the smallest positive principal angle between $L$ and $M$ (defined as $0$ if all the principal angles are $0$), $\PP_M$ denotes the orthogonal projection onto a linear space $M$, and $
	\alpha(\theta)={{2(1+\cos \theta)}/{(1-\cos\theta)^3}}.$
Here, the norm $\|\cdot\|$ could be any unitary invariant, uniformly generated and normalized matrix norm.
In particular, if we take $\|\cdot\|$ to be the spectral norm $\|\cdot\|_2$, then we have
\begin{equation}
\begin{aligned}
	&\sin\angle(L'\cap M',L\cap M)\\
\leq & 8 \max\{
      \alpha(\theta_{\min,+}(L,M)),\alpha(\theta_{\min,+}( L', M'))
      \}(\sin\angle(L,L')+\sin\angle({M},{M'})).
      \end{aligned}
\end{equation}
\end{proposition}
We refer the readers to \cite{stewart1990matrix} for more details on principal angles between two linear spaces and on matrix norms. In particular, the spectral norm is a unitary invariant, uniformly generated and normalized matrix norm.
The result in Proposition~\ref{prop:perturb} holds for all linear subspaces.
The right-hand side in \eqref{eq:perturbation-bound} is finite if and only if $\theta_{\max,+}(L,M)\neq 0$ and $\theta_{\max,+}(L',M')\neq0$.
In our problem, $L=\Col(\Theta_{[1:N,S_1]})$ and $M=\Col(\Theta_{[1:N,S_2]})$ for $S_1,S_2\subset \{1,...,K\}$ and $S_1\neq S_2$. The next lemma further bounds
$\alpha(\theta_{\min,+}(L,M))$ when $L$ and $M$ are column spaces of a matrix,  which is a key step in proving \eqref{eq:error-bound-dim-k}.
\begin{lemma}\label{lemma:angle-lower}
Let $W\in \Real^{N\times K}\setminus\{\mathbf{0}\}$ for some positive integer $N$ and $K$, and $S_1,S_2\subset\{1,...,K\}$ be such that $S_1\setminus S_2\neq \emptyset$ and $S_2\setminus S_1 \neq \emptyset$, then
\begin{equation}\label{eq:angle-lower}
		\cos (\theta_{\min,+}(\Col(W_{[S_1]}),\Col(W_{[S_2]})))\leq 1-\frac{\sigma_{|S_1\cup S_2|}^2(W_{[S_1\cup S_2]})}{\|W\|_2^2}.
	\end{equation}
\end{lemma}

\section{Numerical Experiments}\label{sec:sim}

\subsection{Simulation Study I} We first verify Theorem~\ref{thm:consistency} and its implications when all latent factors are structurally identifiable. Specifically, we consider $K = 5$ under the three models discussed in Section~\ref{subsec:model}, including the linear, the MIRT and the Poisson models. Two design structures are considered, including (1) a simple structure, where $p_Q(\{k\}) = 1/5$, $k = 1, ..., 5$ and (2) a mixed structure, where $p_Q(S)= 1/5$, $S = \{1, 2, 3\}, \{2, 3, 4\}, \{3, 4, 5\}, \{4, 5, 1\}$, and $\{5, 1, 2\}$. The true person parameters
$\ttt_i^*$s and the true manifest parameters
$\aaaa_j^*$s
are generated i.i.d. from distributions over the ball $\{\xx \in \Real^{K}: \Vert \xx\Vert \leq 2.5\}$, respectively, (i.e., $C = 2.5$ in $S_Q$). Under these settings, all the latent factors are structurally identifiable.

For each model and each design structure, a range of $J$ values are considered and we let $N  = 25J$. Specifically, we consider $J = 100, 200, ..., 1000$ for the linear, the MIRT, and the Poisson models.  For each combination of a model, a design structure, and a $J$ value, 50 independent datasets are generated. For each dataset, we apply Algorithm~\ref{alg:jmle} to solve \eqref{eq:mle}, where $C' = 1.2C$.

\begin{figure}
\centering
\begin{subfigure}{.33\textwidth}
  \centering
  \includegraphics[width=1\linewidth]{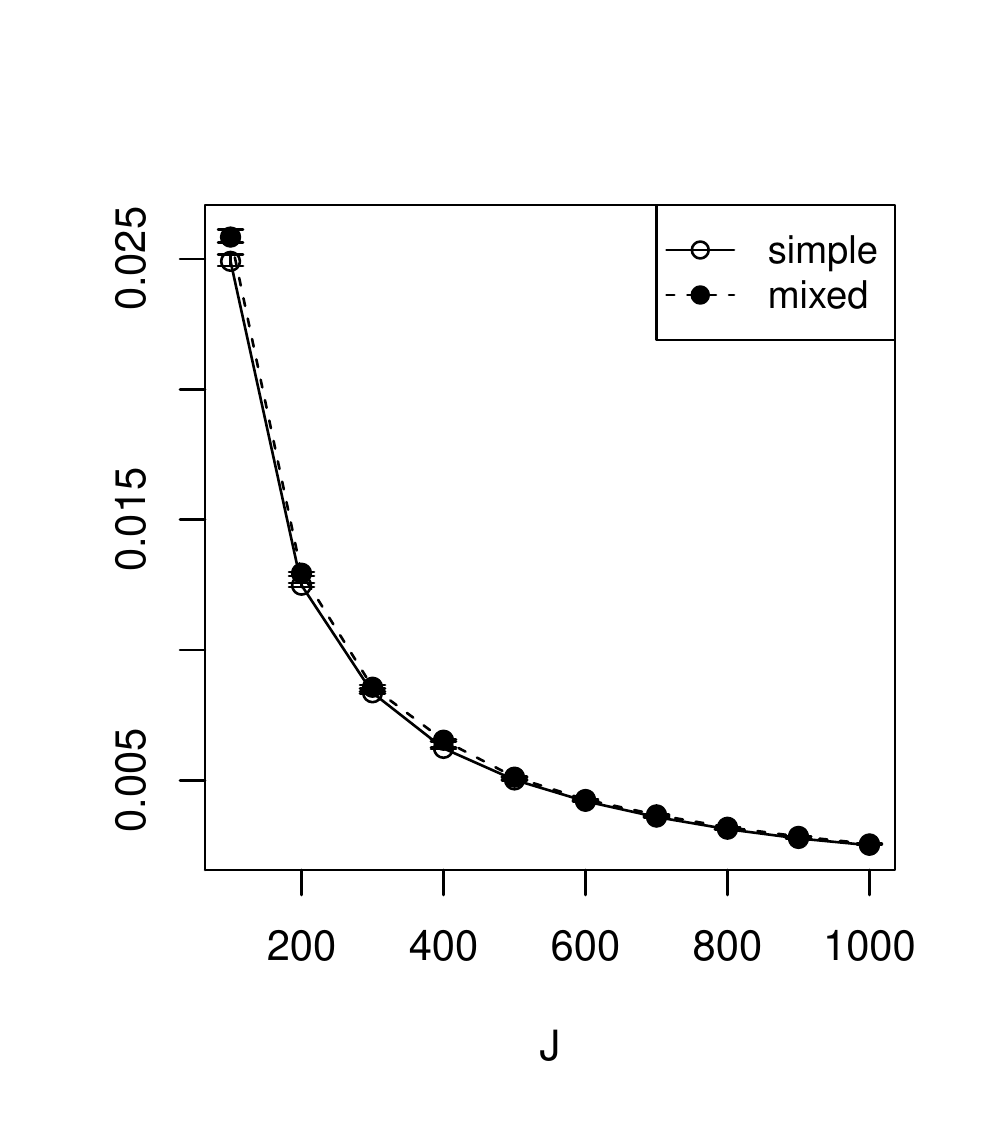}
  \caption{Linear Factor Model}
\end{subfigure}%
\begin{subfigure}{.33\textwidth}
  \centering
  \includegraphics[width=1\linewidth]{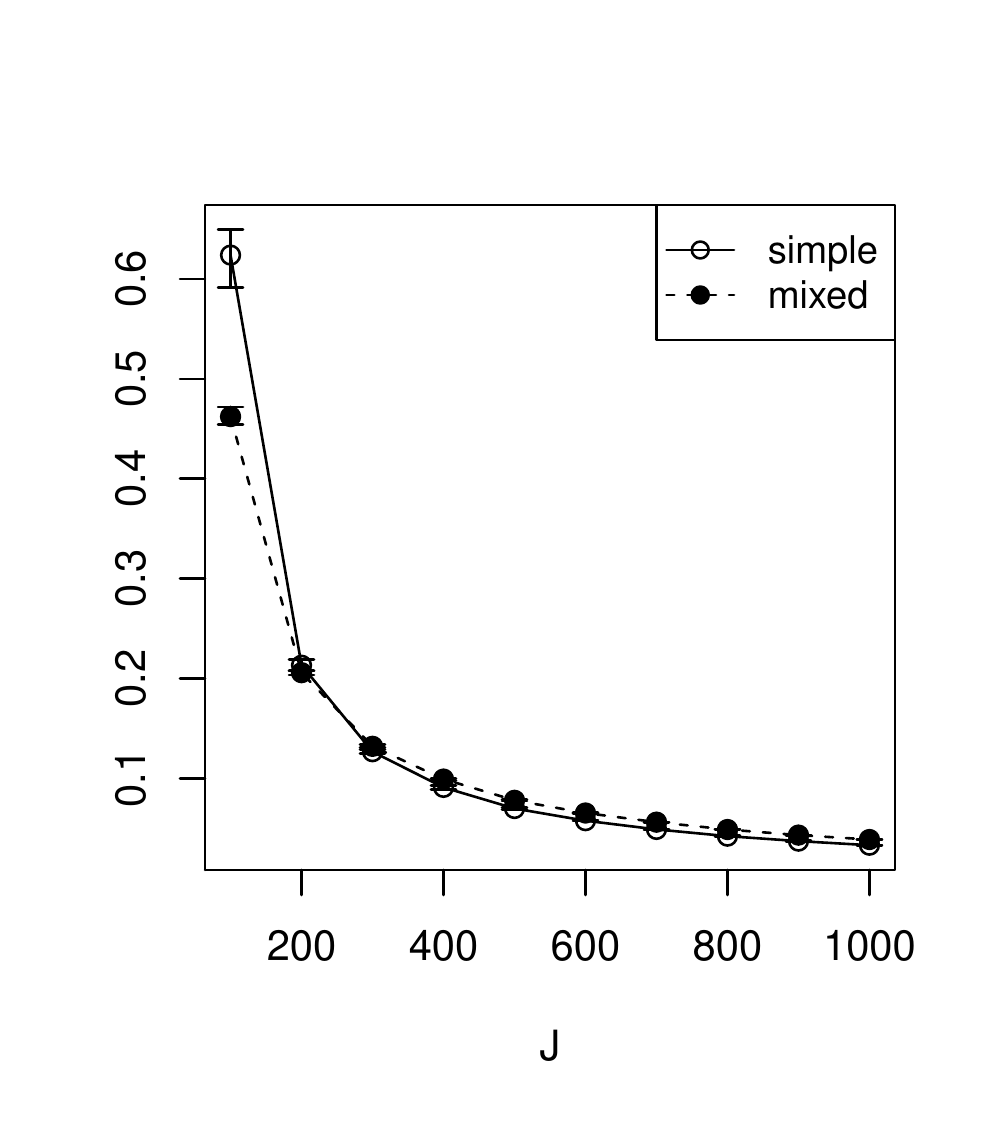}
  \caption{MIRT Model}
\end{subfigure}
\begin{subfigure}{.33\textwidth}
\centering
\includegraphics[width=1\linewidth]{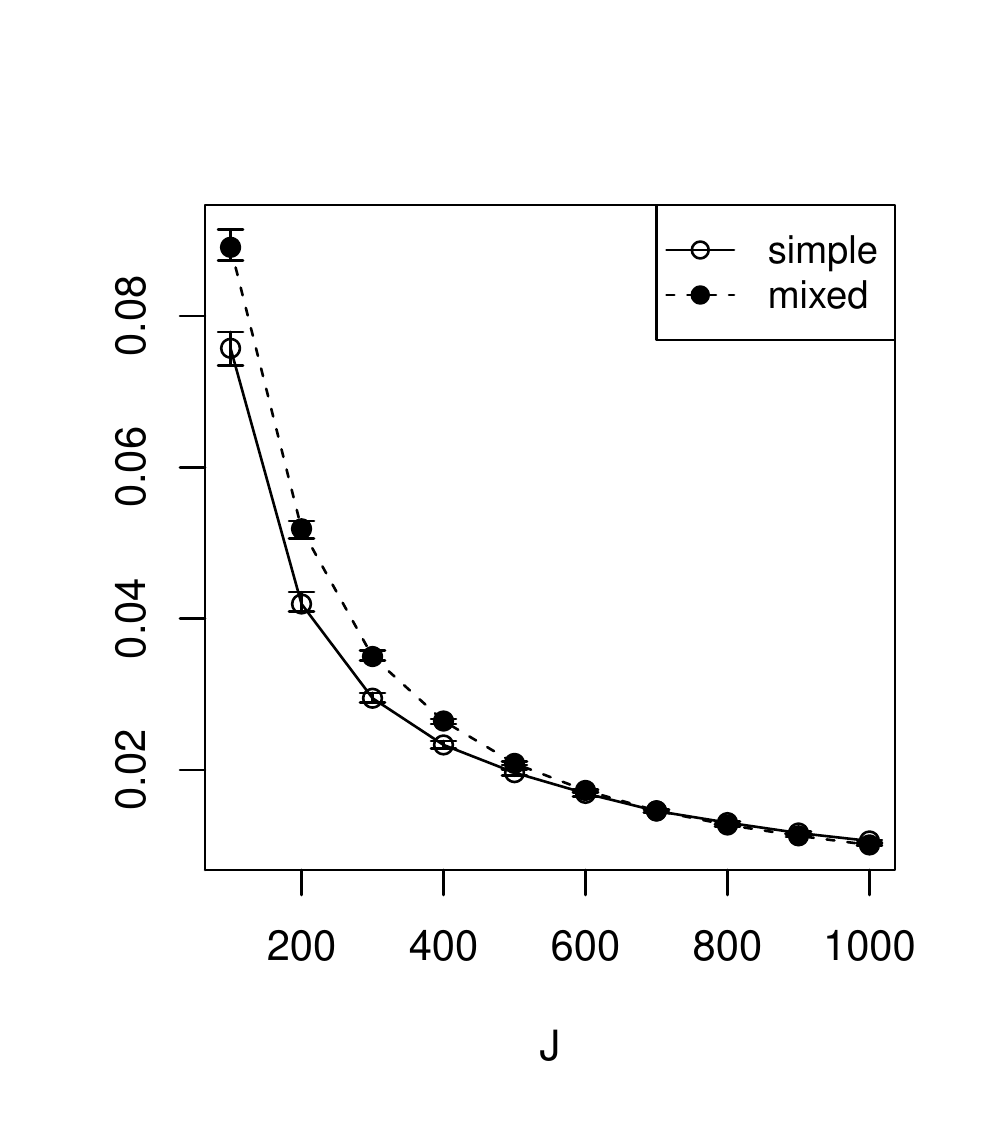}
\caption{Poisson Factor Model}
\end{subfigure}
    \caption{The value of $\|\hat{\Theta}^{(N, J)}(\hat{A}^{(N, J)})^{\top}-\Theta^*_{[1:N, 1:K]}
    A^{*\top}_{[1:J, 1:K]}\|_F^2/(NJ)$ versus the number of manifest variables $J$ under different simulation settings (solid line: simple structure; dashed line: mixed structure). The median, 25\% quantile and 75\% quantile based on the 50 independent replications are shown by the dot, lower bar, and upper bar, respectively.}\label{fig:1}
\end{figure}

\begin{figure}
\centering
\begin{subfigure}{.33\textwidth}
  \centering
  \includegraphics[width=1\linewidth]{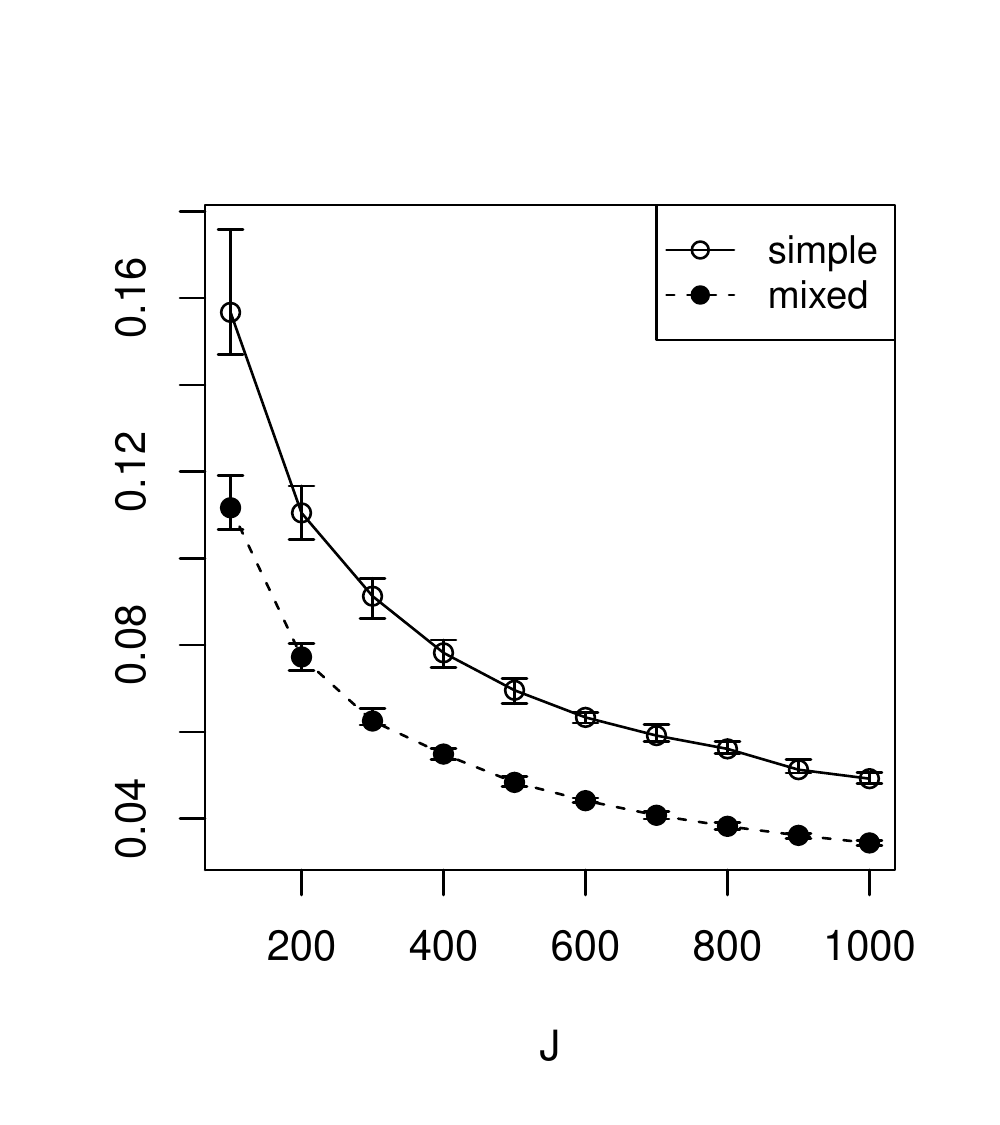}
  \caption{Linear Factor Model}
\end{subfigure}%
\begin{subfigure}{.33\textwidth}
  \centering
  \includegraphics[width=1\linewidth]{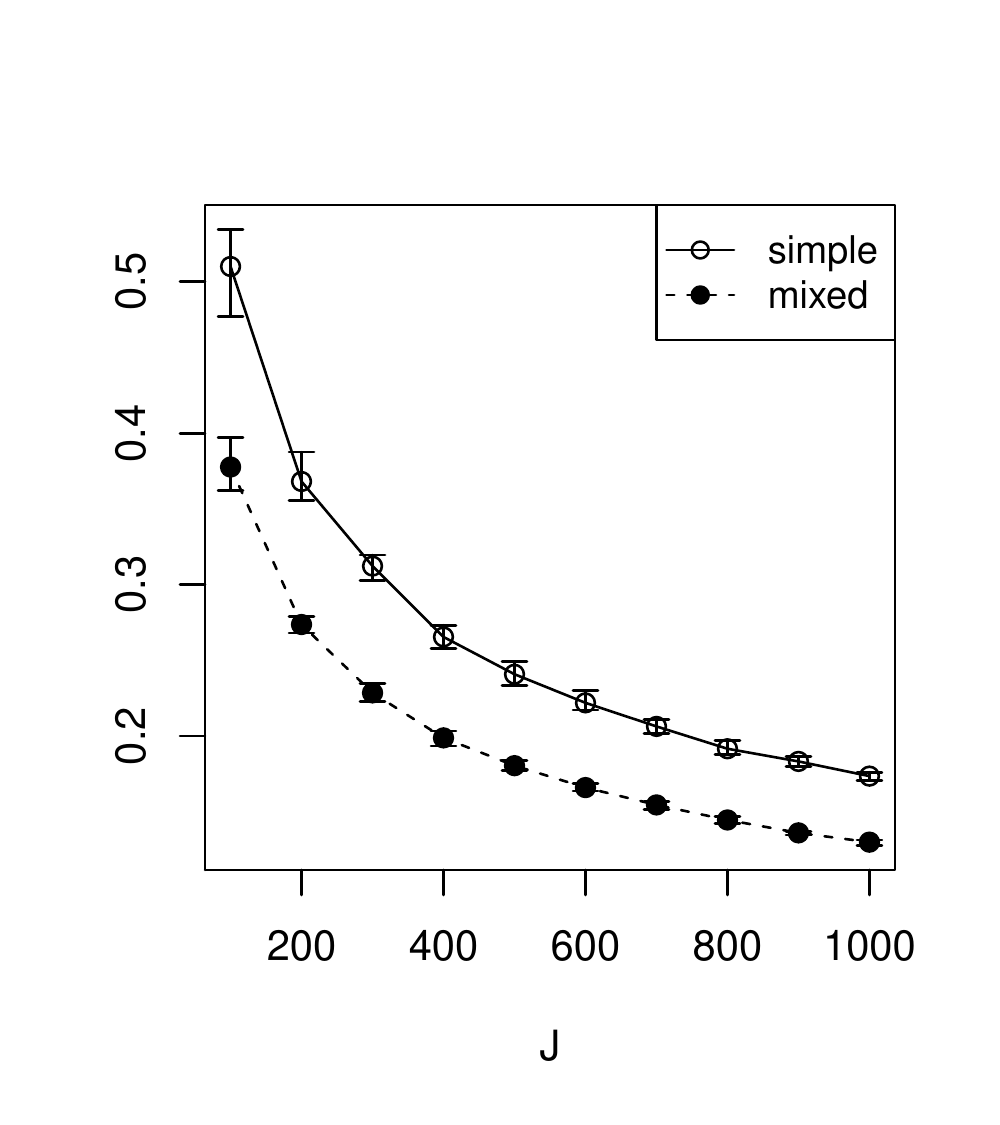}
  \caption{MIRT Model}
\end{subfigure}
\begin{subfigure}{.33\textwidth}
\centering
\includegraphics[width=1\linewidth]{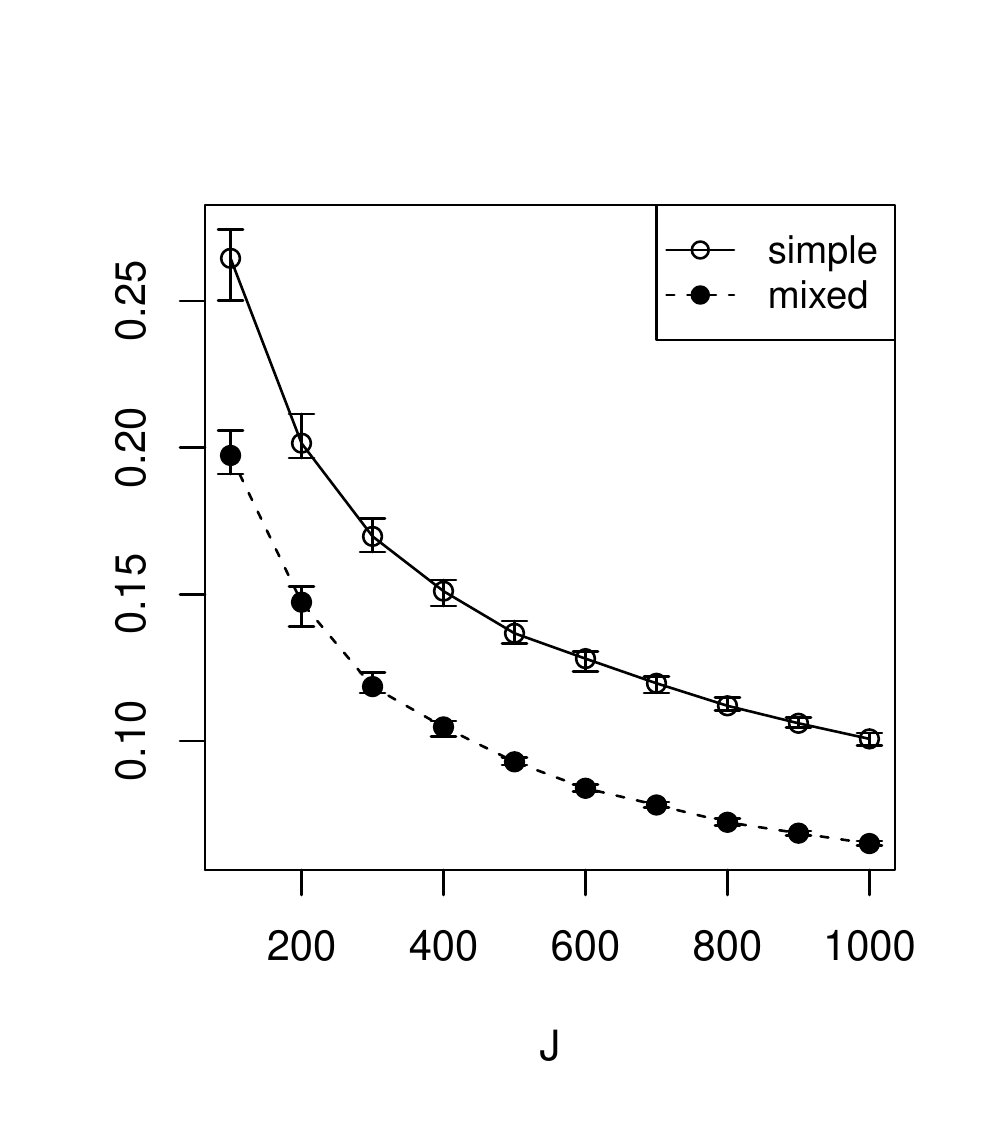}
\caption{Poisson Factor Model}
\end{subfigure}
    \caption{The value of $ \sin\angle(\Theta^*_{[1:N, 1]},\hat{\Theta}_{[1]}^{(N, J)})$  under different simulation settings (solid line: simple structure; dashed line: mixed structure). The median, 25\% quantile and 75\% quantile based on the 50 independent replications are shown by the dot, lower bar, and upper bar, respectively.}\label{fig:2}
\end{figure}

\begin{figure}
\centering
\centering
\begin{subfigure}{.35\textwidth}
  \centering
  \includegraphics[width=1\linewidth]{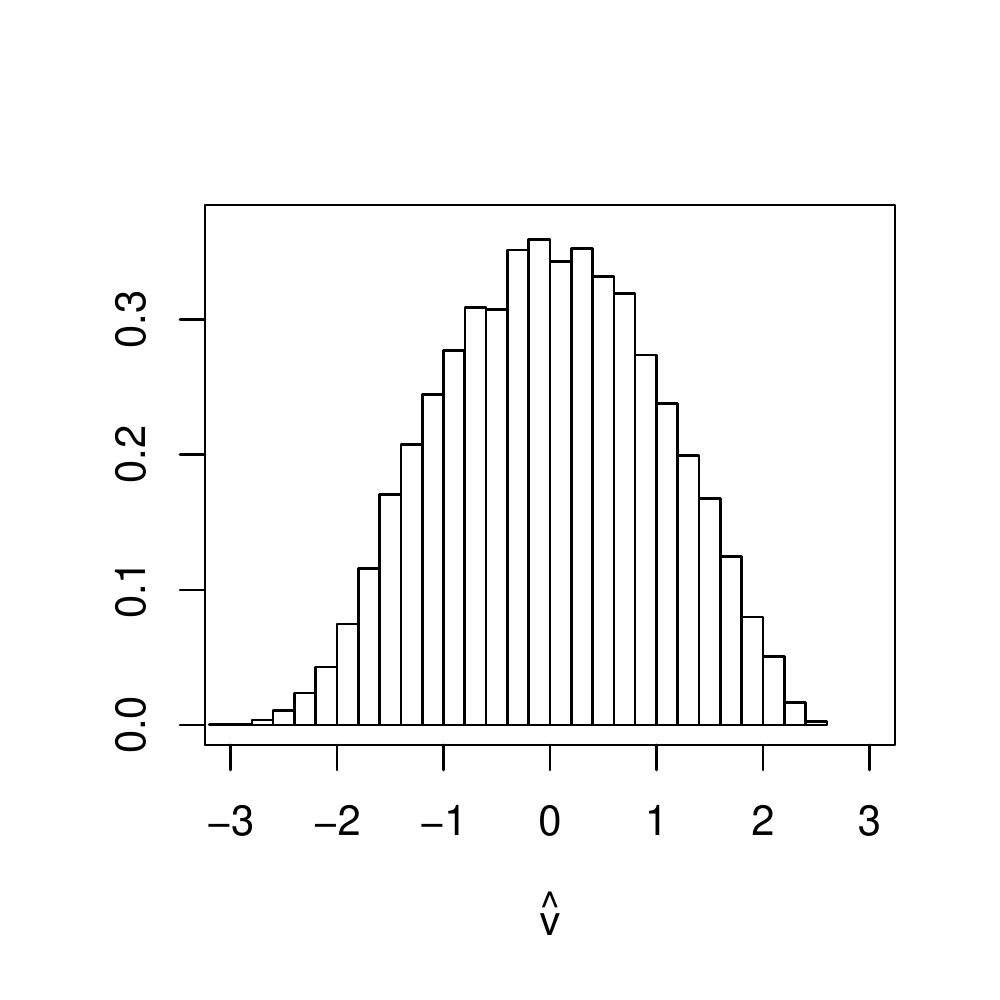}
  \caption{Estimated}
\end{subfigure}%
\begin{subfigure}{.35\textwidth}
  \centering
  \includegraphics[width=1\linewidth]{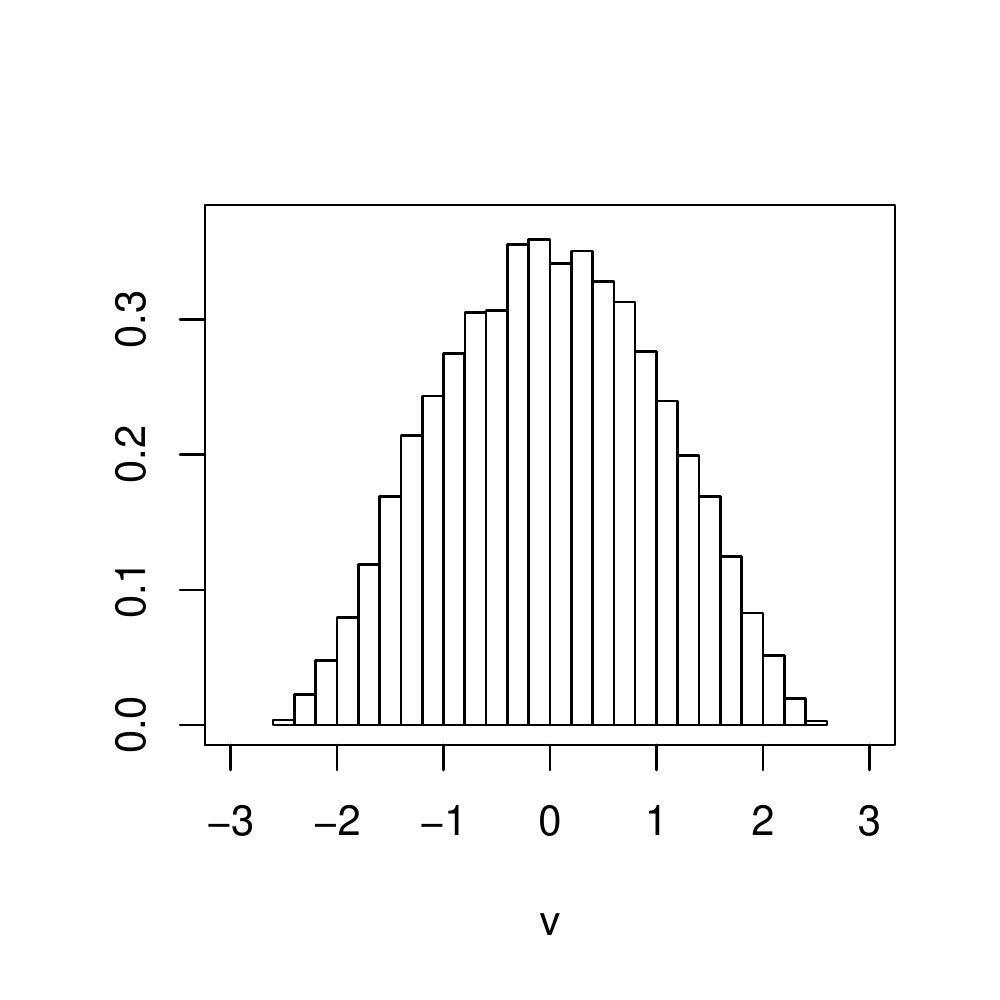}
  \caption{True}
\end{subfigure}
    \caption{Comparison between the histogram of $v_i$s and that of $\hat v_i$s for the first latent factor under the Poisson Model and the simple structure.}\label{fig:3}
\end{figure}

\begin{figure}
\centering
\begin{subfigure}{.33\textwidth}
  \centering
  \includegraphics[width=1\linewidth]{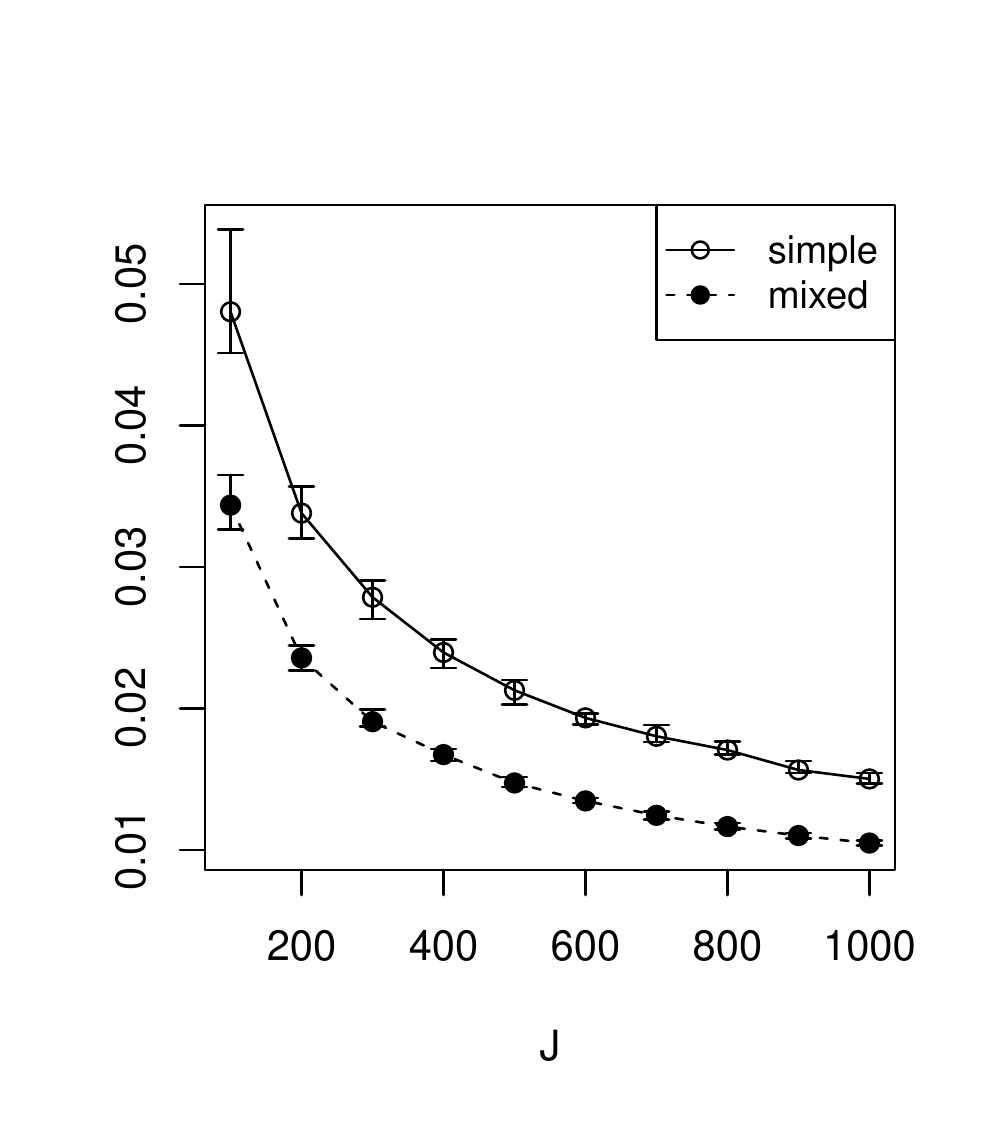}
  \caption{Linear Factor Model}
\end{subfigure}%
\begin{subfigure}{.33\textwidth}
  \centering
  \includegraphics[width=1\linewidth]{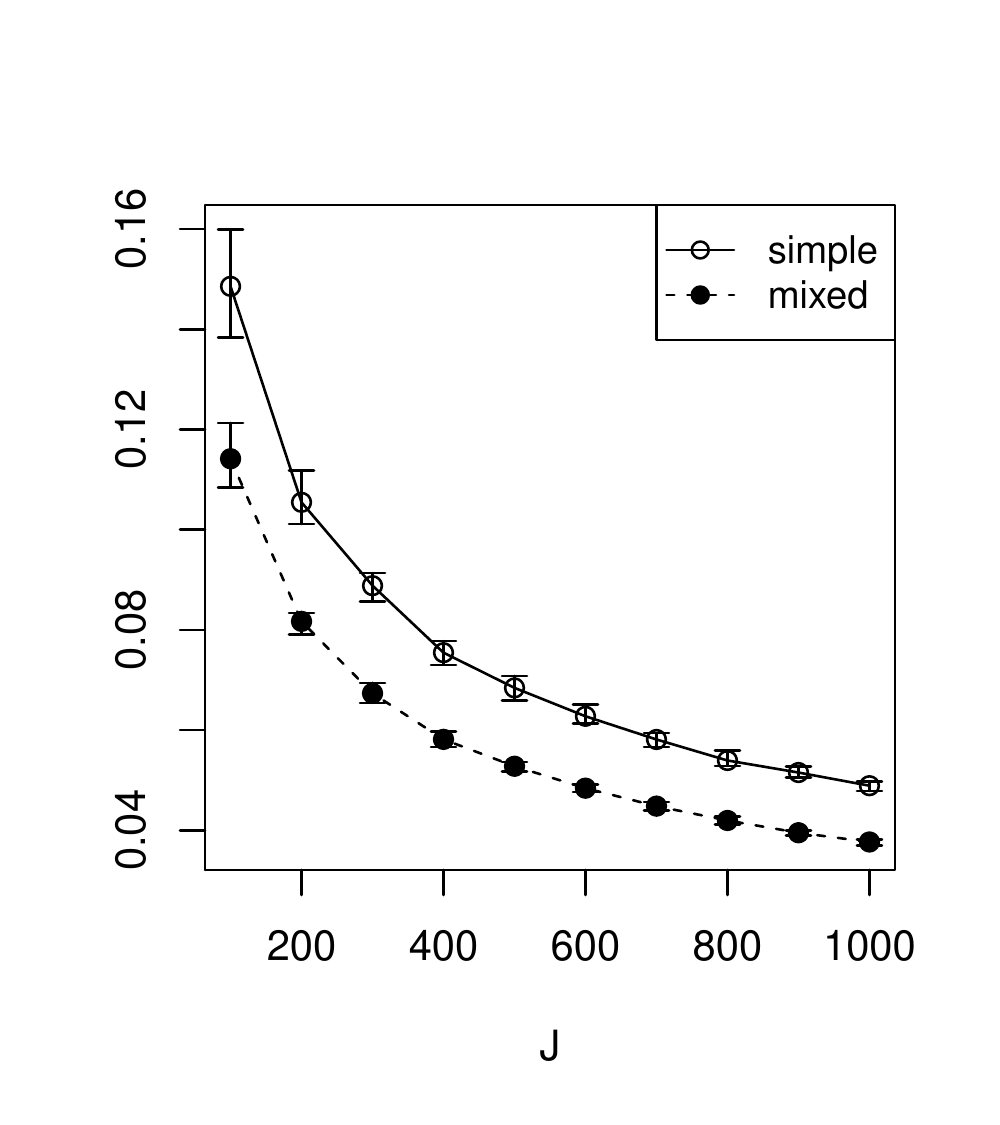}
  \caption{MIRT Model}
\end{subfigure}
\begin{subfigure}{.33\textwidth}
\centering
\includegraphics[width=1\linewidth]{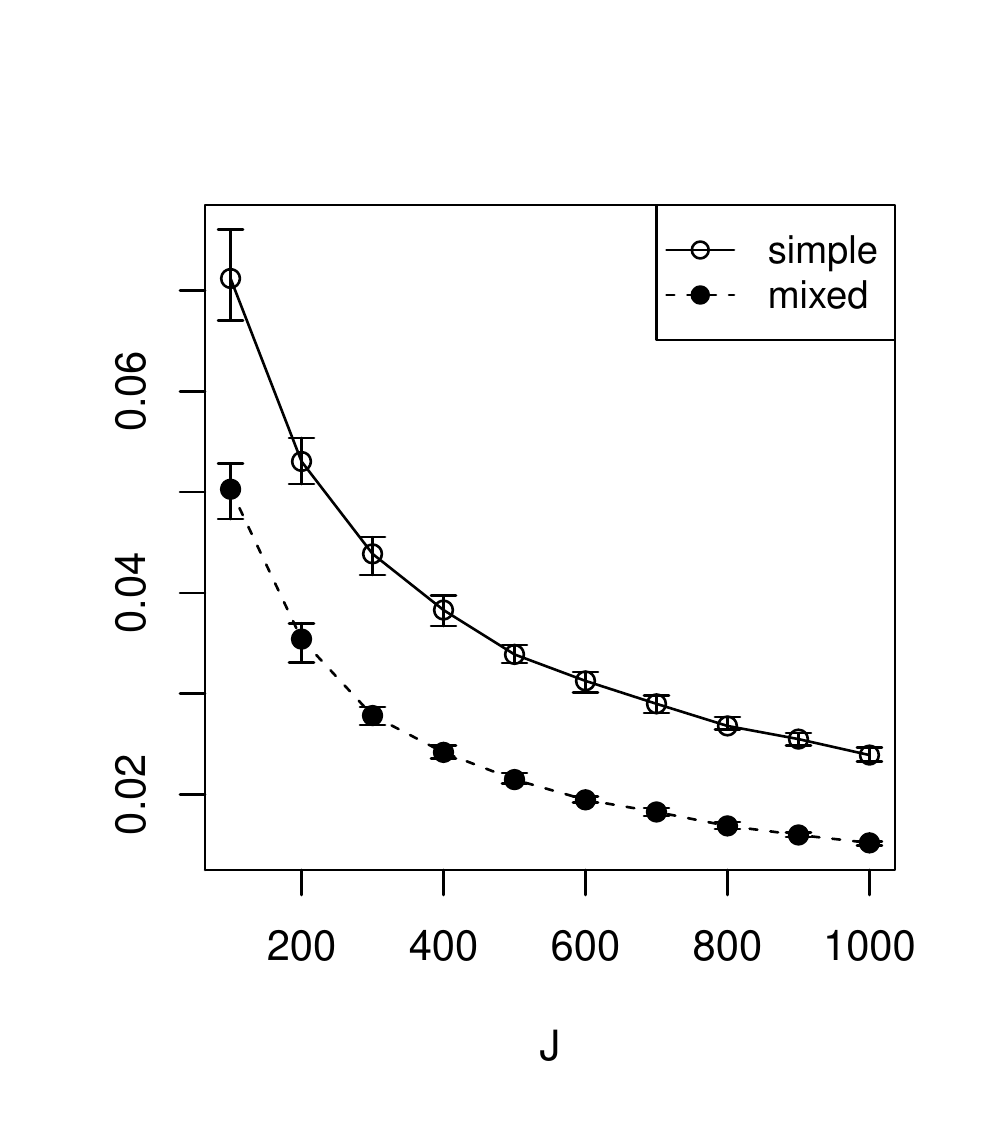}
\caption{Poisson Factor Model}
\end{subfigure}
    \caption{The Kendall's tau ranking error calculated from $\Theta^*_{[1:N, 1]}$ and $\hat{\Theta}_{[1]}^{(N, J)}$,
    under different simulation settings (solid line: simple structure; dashed line: mixed structure). The median, 25\% quantile and 75\% quantile based on the 50 independent replications are shown by the dot, lower bar, and upper bar, respectively.}\label{fig:4}
\end{figure}

\begin{figure}
\centering
\begin{subfigure}{.33\textwidth}
  \centering
  \includegraphics[width=1\linewidth]{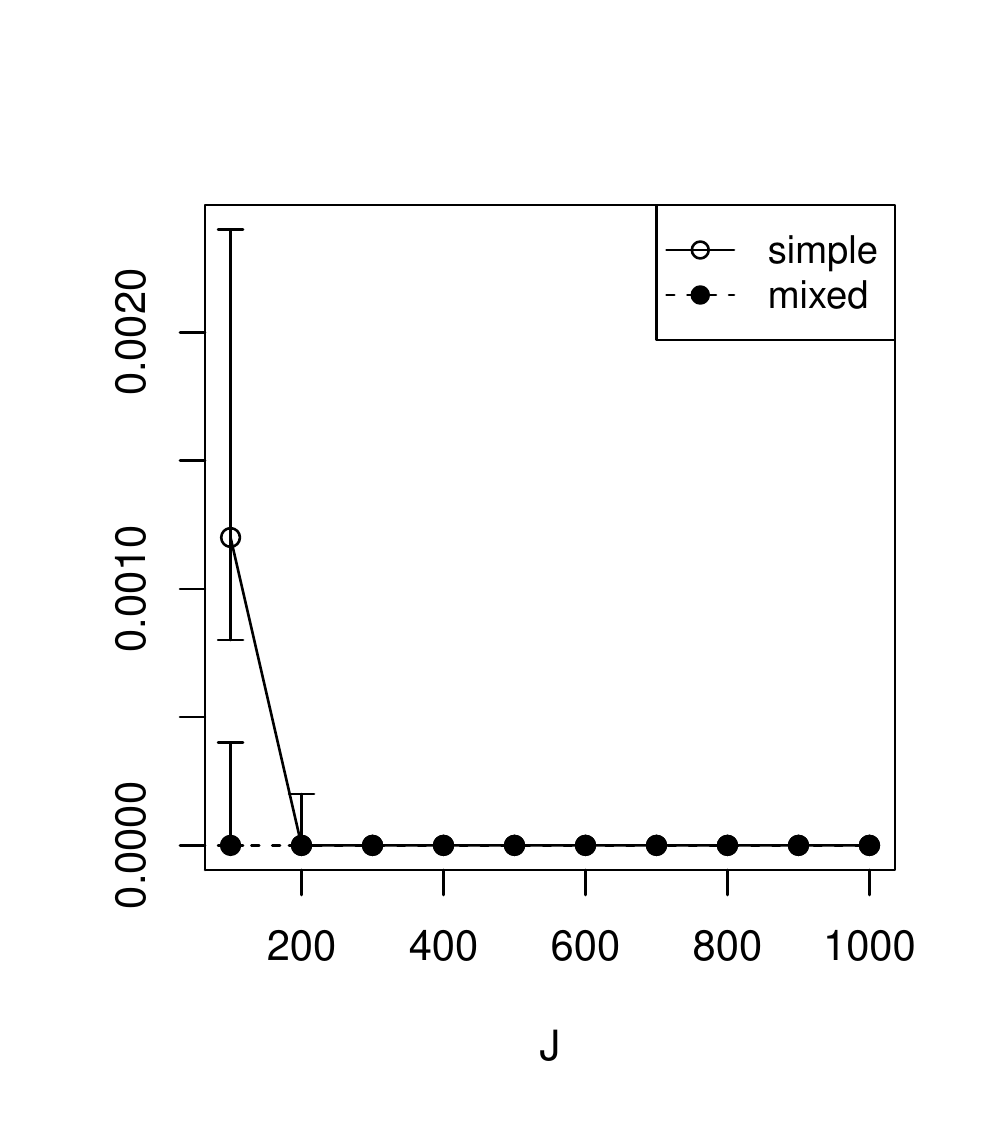}
  \caption{Linear Factor Model}
\end{subfigure}%
\begin{subfigure}{.33\textwidth}
  \centering
  \includegraphics[width=1\linewidth]{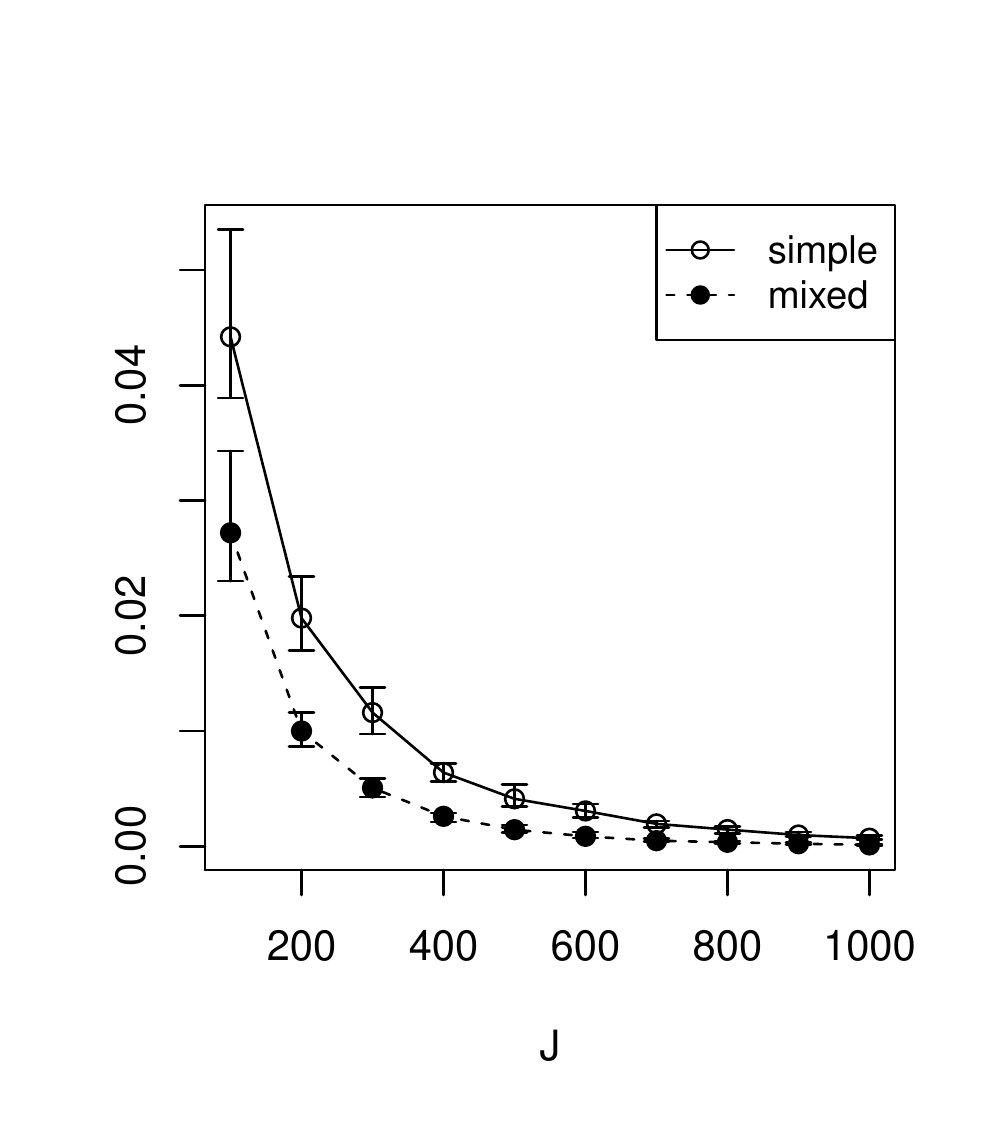}
  \caption{MIRT Model}
\end{subfigure}
\begin{subfigure}{.33\textwidth}
\centering
\includegraphics[width=1\linewidth]{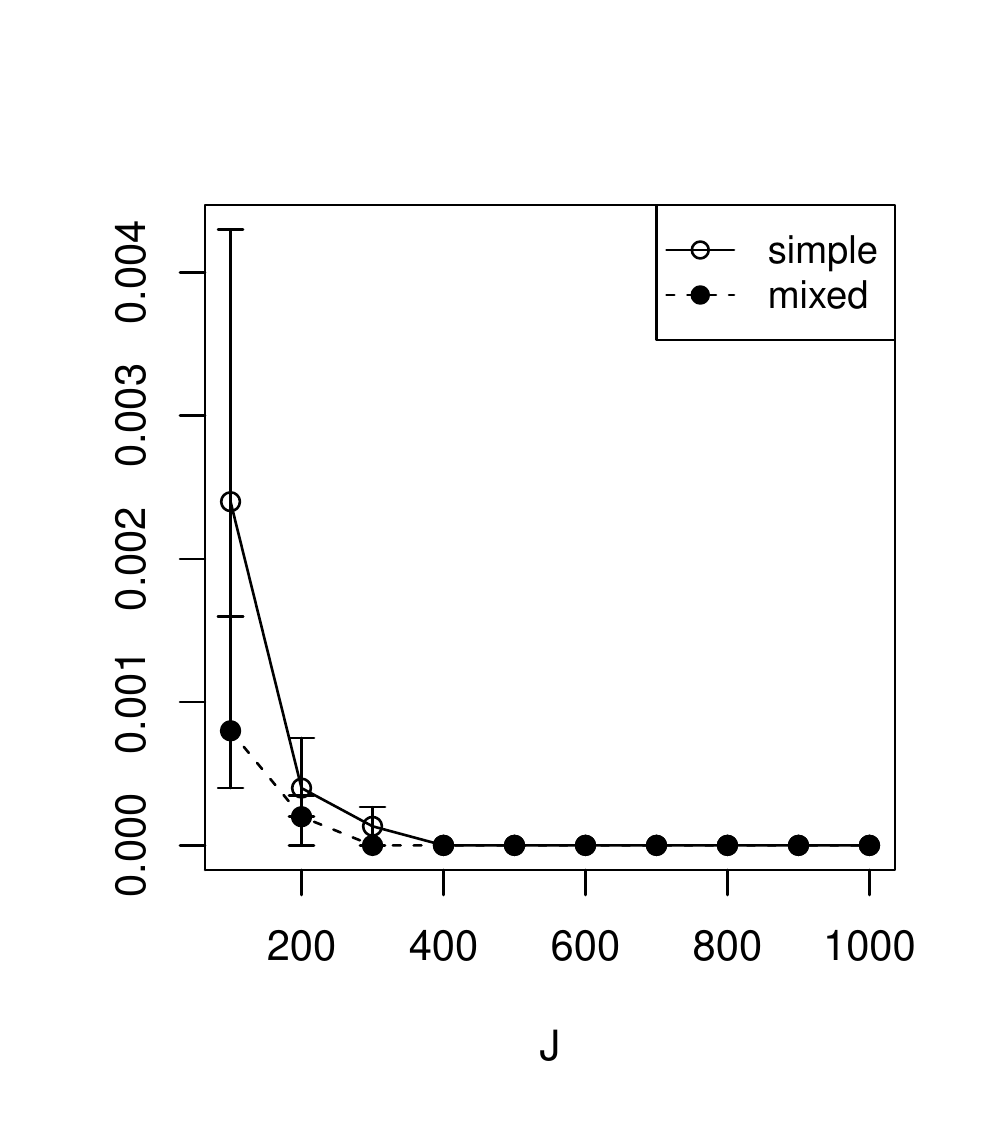}
\caption{Poisson Factor Model}
\end{subfigure}
    \caption{The classification error calculated from $\Theta^*_{[1:N, 1]}$ and $\hat{\Theta}_{[1]}^{(N, J)}$ with indifference zone $(0.13, 0.43)$
    under different simulation settings (solid line: simple structure; dashed line: mixed structure). The median, 25\% quantile and 75\% quantile based on the 50 independent replications are shown by the dot, lower bar, and upper bar, respectively.}\label{fig:5}
\end{figure}

Results are shown in Figures~\ref{fig:1}-\ref{fig:5}.
Figure~\ref{fig:1} shows the trend of $\frac{1}{NJ}\|\hat{\Theta}^{(N, J)}(\hat{A}^{(N, J)})^{\top}-\Theta^*_{[1:N, 1:K]} A^{*\top}_{[1:J, 1:K]}\|_F^2$ ($y$-axis) when $J$ increases ($x$-axis), where each panel corresponds to a model. This result verifies \eqref{eq:error-bound-m} in Theorem~\ref{thm:consistency}.
According to these plots, the normalized squared Frobenius norm between
$\Theta^*_{[1:N, 1:K]} A^{*\top}_{[1:J, 1:K]}$ and its estimate $\hat{\Theta}^{(N, J)}(\hat{A}^{(N, J)})^{\top}$ decays towards zero, as $J$ and $N$ increase.
Figures~\ref{fig:2}-\ref{fig:5} present results based on the first latent factor and we point out that
the results based on the other latent factors are almost the same.
Figure~\ref{fig:2} is used to verify \eqref{eq:error-bound-dim-k}
in Theorem~\ref{thm:consistency}, showing the pattern that $\sin\angle(\Theta^*_{[1:N, k]},\hat{\Theta}_{[k]}^{(N, J)})$ decreases as $J$ and $N$ increase.

Moreover, Figure~\ref{fig:3} provides evidence on the result of Proposition~\ref{prop:empdist}.
Displayed in Figure~\ref{fig:3} are the histograms of $v_i$s and $\hat v_i$s, respectively,
based on a randomly selected dataset when $J = 1000$ under the Poisson model and the simple structure.
According to this figure, little difference is observed between the empirical distribution of $v_i$s and that of $\hat v_i$s. Similar results are observed for other datasets under all these three models when $J$ and $N$ are large.

Finally,
Figures~\ref{fig:4} and \ref{fig:5} show results of ranking and classification based on the estimated factor scores, whose theoretical results are given in Propositions~\ref{prop:ranking} and \ref{prop:classify}.
The $y$-axes of the two figures show the normalized Kendall's tau distance in \eqref{eq:ranking}
and the classification error in \eqref{eq:classificatin}, respectively.
Specifically, $\tau_-$ and $\tau_+$ are chosen as
0.14 and 0.43 which are the $55\%$ and $65\%$ quantiles of the $v_i$s.
From these plots, both the ranking and the classification errors tend to zero, as $J$ and $N$ grow large.

\subsection{Simulation Study II}
We then provide an example, in which a latent factor is not identifiable. Specifically, we consider $K = 2$ and the same latent factor models as in Study I.
The design structure is given by $p_Q(\{1\}) = 1/2$ and $p_Q(\{1, 2\}) = 1/2$.
The true person parameters
$\ttt_i^*$s and the true manifest parameters
$\aaaa_j^*$s
are generated i.i.d. from distributions over the ball $\{\xx \in \Real^{K}: \Vert \xx\Vert \leq 3\}$, respectively, (i.e., $C = 3$ in $S_Q$). Under these settings, the first latent factor is structurally identifiable and the second factor is not.
For each model, we consider $J = 100, 200, ..., 1000$. The rest of the simulation setting is the same as Study I.
\begin{figure}
\centering

\includegraphics[width=.33\linewidth]{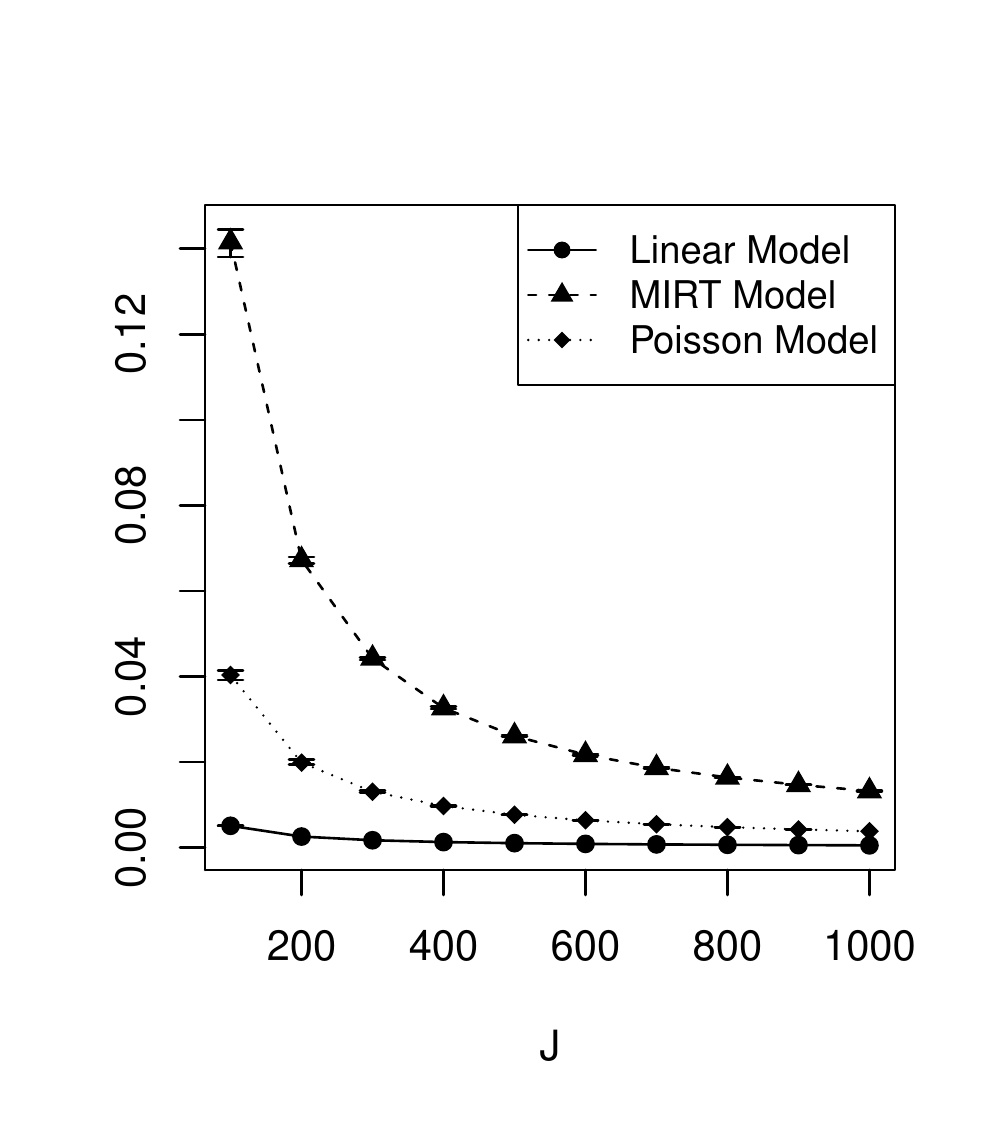}

\caption{The value of $\|\hat{\Theta}^{(N, J)}(\hat{A}^{(N, J)})^{\top}-\Theta^*_{[1:N, 1:K]}
    A^{*\top}_{[1:J, 1:K]}\|_F^2/(NJ)$ versus the number of manifest variables $J$ under different simulation settings (solid line: Linear Model; dashed line: MIRT Model; dotted line: Poisson Model). The median, 25\% quantile and 75\% quantile based on the 50 independent replications are shown by the dot, lower bar, and upper bar, respectively.}\label{fig:6}
\end{figure}
\begin{figure}
\centering
\begin{subfigure}{.33\textwidth}
  \centering
  \includegraphics[width=1\linewidth]{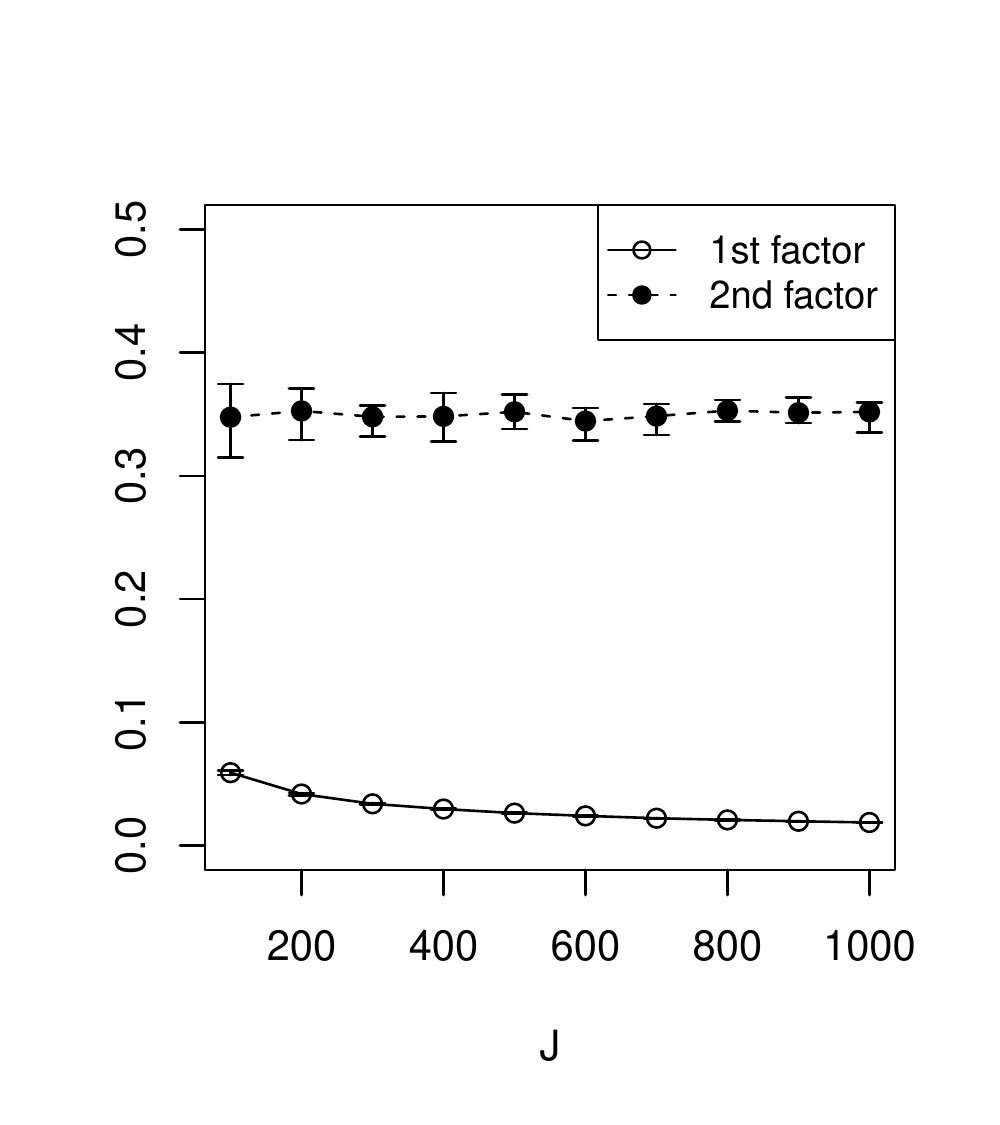}
  \caption{Linear Factor Model}
\end{subfigure}%
\begin{subfigure}{.33\textwidth}
  \centering
  \includegraphics[width=1\linewidth]{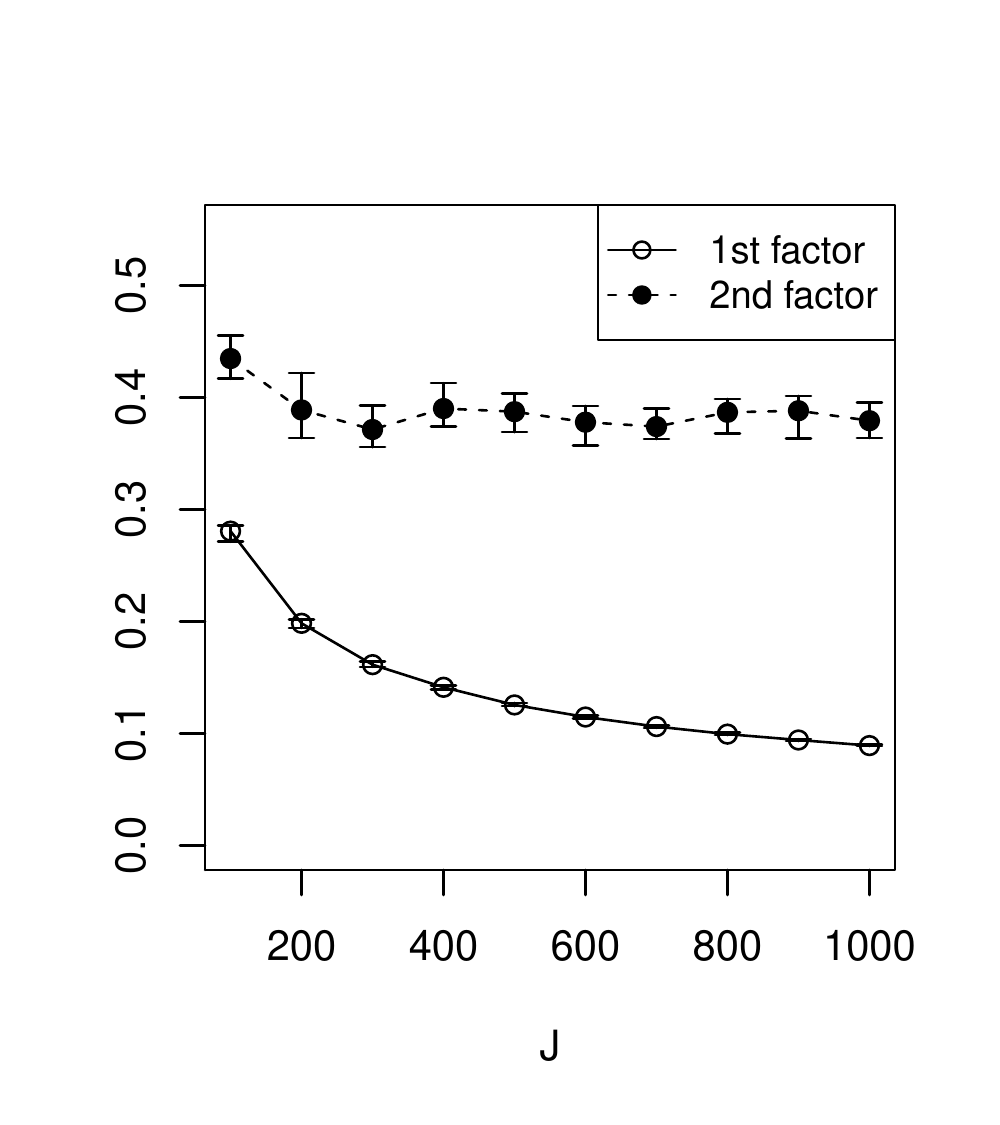}
  \caption{MIRT Model}
\end{subfigure}
\begin{subfigure}{.33\textwidth}
\centering
\includegraphics[width=1\linewidth]{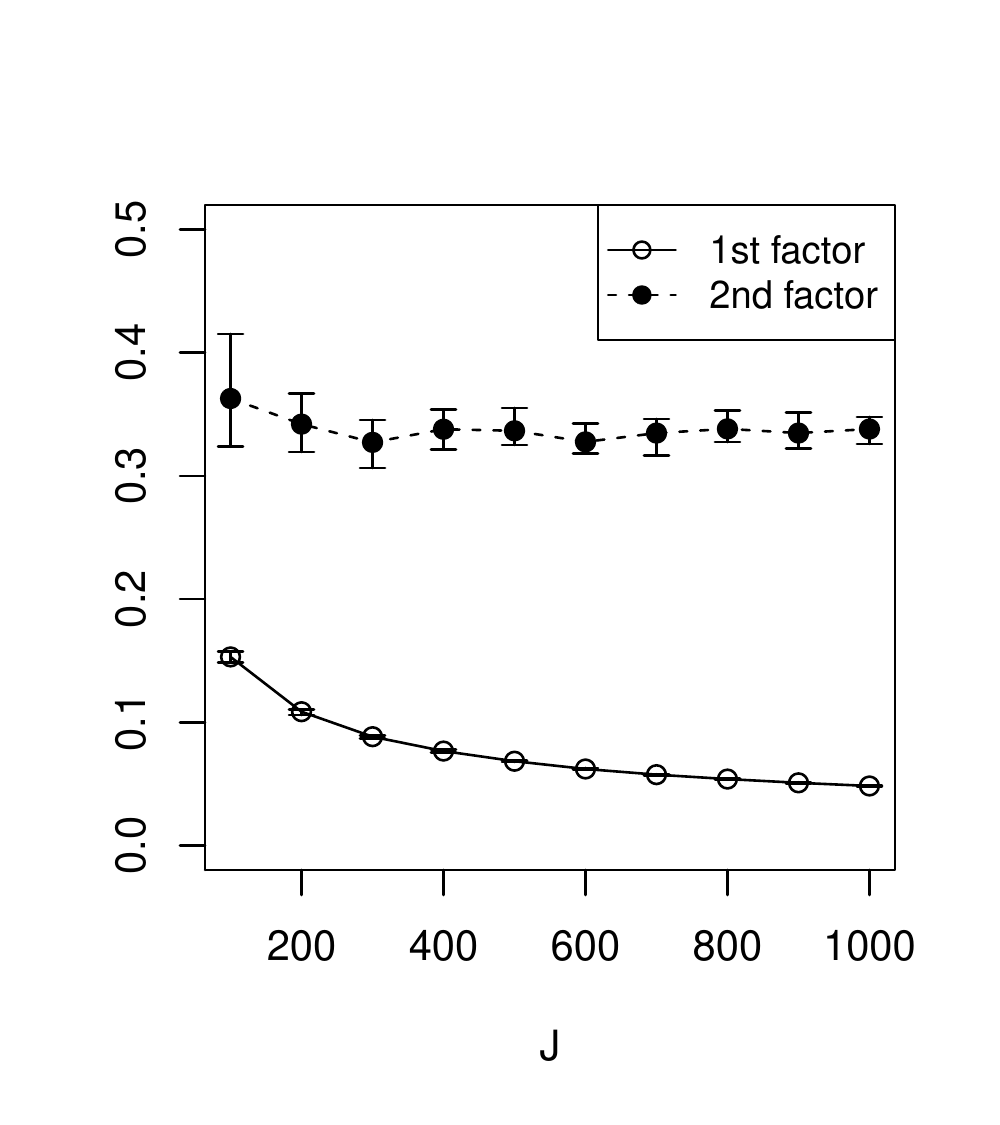}
\caption{Poisson Factor Model}
\end{subfigure}
    \caption{The value of $ \sin\angle(\Theta^*_{[1:N, 1]},\hat{\Theta}_{[1]}^{(N, J)})$  under different simulation settings (solid line: the first latent trait; dashed line: the second latent trait). The median, 25\% quantile and 75\% quantile based on the 50 independent replications are shown by the dot, lower bar, and upper bar, respectively.}\label{fig:7}
\end{figure}
Results are shown in Figures~\ref{fig:6} and \ref{fig:7}. First, Figure~\ref{fig:6} presents the patten that $\frac{1}{NJ}\|\hat{\Theta}^{(N, J)}(\hat{A}^{(N, J)})^{\top}-\Theta^*_{[1:N, 1:K]} A^{*\top}_{[1:J, 1:K]}\|_F^2$
decays to 0 when $J$ increases, even when a latent factor is not structurally identifiable. This is consistent with the first part of Theorem~\ref{thm:consistency}. Second, Figure~\ref{fig:7} shows the
trend of $ \sin\angle(\Theta^*_{[1:N, k]},\hat{\Theta}_{[k]}^{(N, J)})$ as $J$ increases. In particular,
the value of $ \sin\angle(\Theta^*_{[1:N, k]},\hat{\Theta}_{[k]}^{(N, J)})$ stays above 0.3 for most of the data sets for the factor which is structurally unidentifiable, while it still decays towards 0 for the identifiable one.

\subsection{Real Data Example}\label{sec:real}

As an illustration, we apply the proposed method to analyze a personality assessment dataset based on an International Personality Item Pool (IPIP) NEO personality inventory \citep{johnson2014measuring}. This inventory is a public-domain version of the widely used NEO personality inventory \citep{costa1985neo}, which is designed to measure the Big Five personality factors ($K = 5$), including Neuroticism (N), Agreeableness (A), Extraversion (E), Openness to experience (O),  and  Conscientiousness (C).
The dataset contains 20,993 individuals and 300 items. We use a subset of the dataset which contains 7,325 individuals who have answered all 300 items. The measurement design matrix has a simple structure, which is a safe design according to our identifiability theory. Under this design, each item only measures one personality factor and each factor is measured by 60 items.
All the items are on a five-category rating scale, where reverse-worded items were reversely recorded
(1 $\rightarrow$ 5, 2 $\rightarrow$ 4, 4 $\rightarrow$ 2, 5 $\rightarrow$ 1) at the time the respondent completed the inventory.
In this analysis we dichotomize them by merging categories $\{1, 2, 3\}$ and $\{4, 5\}$, respectively, and then fit the MIRT model. 
\begin{figure}
\centering
\includegraphics[width=1\linewidth]{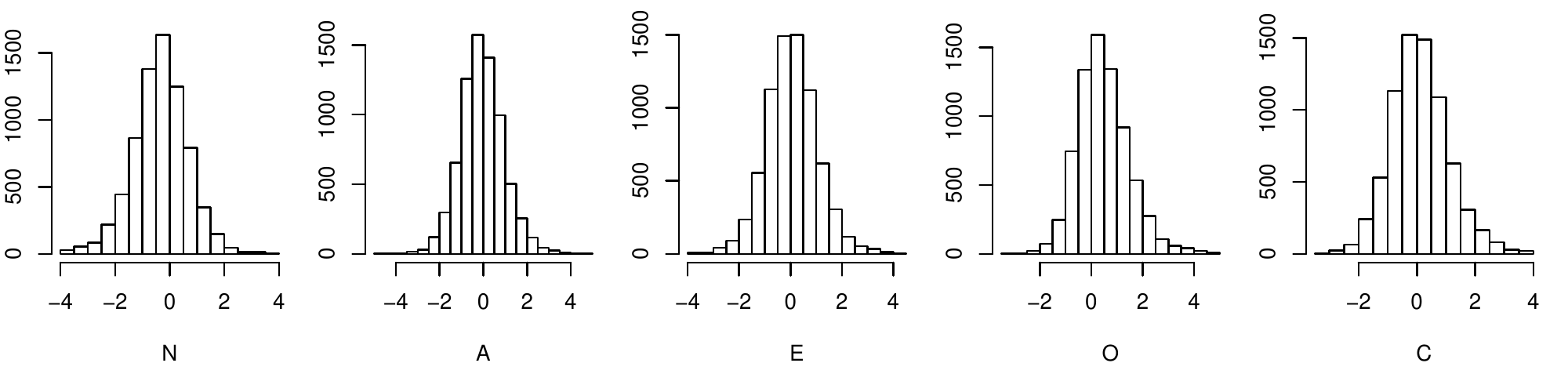}
\caption{The histogram of the estimated factor scores $\hat{\Theta}_{[k]}^{(N, J)},$ $k = 1, ..., 5$ for the IPIP-NEO dataset.}\label{fig:real_hist_theta}
\end{figure}

\begin{table}
\centering
\begin{tabular}{c|ccccc}
\hline
   & N & A & E & O & C \\
 \hline
N  &  1.00 & -0.23 & -0.35 & -0.04 & -0.34\\
A  & -0.23 &  1.00 &  0.22 &  0.19 &  0.31\\
E  & -0.35 &  0.22 &  1.00 &  0.20 &  0.18\\
O  & -0.04 &  0.19 &  0.20 &  1.00 & -0.01\\
C & -0.34 &  0.31 &  0.18 & -0.01 &  1.00\\
\hline
\end{tabular}
\caption{The correlation matrix between the estimated factor scores for the IPIP-NEO dataset.}\label{tab:real_cor_theta}
\end{table}

\begin{figure}
\centering
\includegraphics[width=0.33\linewidth]{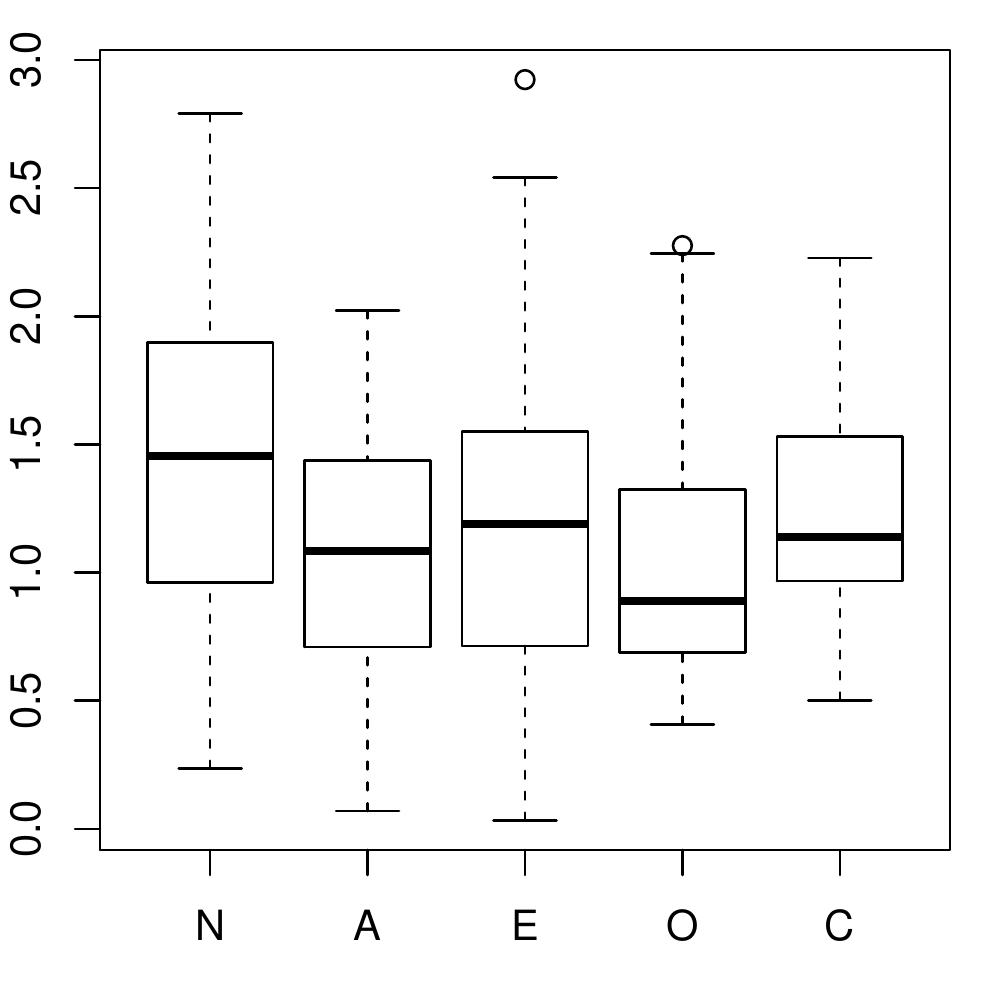}
\caption{The boxplot of the 60 estimated unconstrained loadings for each of the five factors for the IPIP-NEO dataset.}\label{fig:real_boxplot_A}
\end{figure}

\begin{table}
  \centering
  \footnotesize
  \begin{tabular}{c|rcl}
    \hline
    Factor &Items & Loading  & Content \\
     \hline
& 23(N$+$)  & 2.81  & Am often down in the dumps. \\
& 4(N$+$)   & 2.66  & Get stressed out easily. \\
N& 58(N$-$)  & 2.51  & Know how to cope. \\
& 25(N$+$)  & 2.38  & Have frequent mood swings. \\
& 13(N$+$)  & 2.31  & Get upset easily. \\
\hline
& 79(A$-$)  & 2.02  & Take advantage of others. \\
& 86(A$-$)  & 1.88  & Look down on others. \\
A& 84(A$+$)  & 1.82  & Am concerned about others. \\
& 98(A$-$)  & 1.80   & Insult people. \\
& 67(A$-$)  & 1.80   & Distrust people. \\
\hline
& 123(E$+$)     & 2.93  & Feel comfortable around people. \\
& 128(E$-$)     & 2.54  & Avoid contacts with others. \\
E& 124(E$+$)     & 2.48  & Act comfortably with others. \\
& 139(E$-$)     & 2.25  & Avoid crowds. \\
& 132(E$+$)     & 2.19  & Talk to a lot of different people at parties. \\
\hline
& 196(O$-$)     & 2.28  & Do not like art. \\
& 191(O$+$)     & 2.25  & Believe in the importance of art. \\
O& 226(O$-$)     & 2.11  & Am not interested in abstract ideas. \\
& 225(O$+$)     & 2.07  & Enjoy thinking about things. \\
& 229(O$-$)     & 2.05  & Am not interested in theoretical discussions. \\
\hline
& 286(C$-$)     & 2.23  & Find it difficult to get down to work. \\
& 272(C$+$)     & 2.21  & Work hard. \\
C& 283(C$+$)     & 2.15  & Start tasks right away. \\
& 289(C$-$)     & 2.11  & Have difficulty starting tasks. \\
& 285(C$+$)     & 2.10   & Carry out my plans. \\
\hline
\end{tabular}
\caption{The content of the top five items with highest estimated loadings on each factor for the IPIP-NEO dataset.}\label{tab:real_top_5}
\end{table}

The results are shown in Figures~\ref{fig:real_hist_theta} through \ref{fig:real_boxplot_A} and Tables~\ref{tab:real_cor_theta} and Table~\ref{tab:real_top_5}. In Figure~\ref{fig:real_hist_theta}, the histograms of $\hat{\Theta}_{[k]}^{(N, J)}$ are given, for $k = 1, ..., 5$, which correspond to the N, A, E, O, and C factors, respectively.  As we can see, the estimated factor scores are quite normally distributed, especially for the first three factors. For the O and C factors, the distributions of the estimated factor scores are slightly right skewed. In Table~\ref{tab:real_cor_theta}, the correlations between the factors are calculated using the estimated factor scores. The correlations between the factors are relatively small, which  
are largely consistent with the existing findings in the literature of Big Five personality \citep{digman1997higher}. Figure~\ref{fig:real_boxplot_A} shows the box plot of the 60 unconstrained loadings for each of the five factors, according to which all the estimated loadings take value between 0 and 3 and the majority of them take value between 0.5 and 2. Table~\ref{tab:real_top_5} lists the content of the top five items with the highest estimated loadings for each factor. These items are indeed representative of the Big Five factors.


\section{Concluding Remarks}\label{sec:conc}

In this paper, we study how design information affects the identifiability and estimability of a general family of structured latent factor models, through both asymptotic and non-asymptotic analyses.
In particular,  under a double asymptotic regime where both the numbers of individuals and manifest variables grow to infinity, we define the concept of structural identifiability for latent factors.
Then
necessary and sufficient conditions are established for the structural identifiability of a given latent factor. Moreover, an estimator is proposed that can consistently recover all the structurally identifiable latent factors. A non-asymptotic error bound is developed
to characterize the effect of design information on the estimation of latent factors, which complements the asymptotic results. In establishing these results, new perturbation bounds on the intersection of linear subspaces, as well as some other technical tools, are developed, which may be of independent theoretical interest. As shown in Section~\ref{sec:implication},
our results have significant implications on the use of generalized latent factor models in large-scale educational and psychological measurement.

There are many future directions along the current work. First, it is of interest to develop methods, such as information criteria, for model comparison based on the proposed estimator. These methods can be used to
select a design matrix $Q$ that best describes the data structure when there are multiple candidates,
or to determine the underlying latent dimensions.
Second, the current results may be further generalized by considering more general latent factor models beyond the exponential family. For example,
we may establish similar identifiability and estimability results when the distribution of $Y_{ij}$ is a
more complicated function or even an unknown function of $\ttt_i$ and $\aaaa_j$.

\section*{Acknowledgement}
We would like to thank the editors and two referees for their helpful and constructive comments. We would also like to thank Prof. Zhiliang Ying for his valuable comments.
Xiaoou Li's research is partially supported by the NSF grant DMS-1712657. Yunxiao Chen's research is supported in part by NAEd/Spencer postdoctoral fellowship.

\bibliographystyle{apalike}

\end{document}